\documentclass[letterpaper,peerreview,draftclsnofoot,onecolumn,11pt]{IEEEtran}
\usepackage{amsmath, amssymb, graphicx, cite, color}

\usepackage{algorithm,algorithmic,subfigure}

\newtheorem{theorem}{Theorem}
\newtheorem{lemma}{Lemma}
\newtheorem{proposition}{Proposition}
\newtheorem{corollary}{Corollary}
\newtheorem{remark}{Remark}

\newtheorem{definition}{Definition}

\newcommand{\be}{\begin{eqnarray}}
\newcommand{\ee}{\end{eqnarray}}

\newcommand{\mc}{\mathcal}

\IEEEoverridecommandlockouts 

\title{Energy-Reliability Limits in Nanoscale Feedforward Neural Networks and Formulas \thanks{This work was supported in part by Systems on Nanoscale Information fabriCs 
(SONIC), one of the six SRC STARnet Centers, sponsored by MARCO and DARPA.} \thanks{This work was presented in part at the 2016 Information Theory and Applications Workshop \cite{ChatterjeeV2016}, and the 2017 Conference
on Information Science and Systems \cite{ChatterjeeV2017a}.}\thanks{A. Chatterjee is with the Department of  Electrical Engineering, Indian Institute of Technology Madras, Chennai 600036, India (e-mail: avhishek@ee.iitm.ac.in). L.~R. Varshney is with the Coordinated Science Laboratory, University of Illinois at Urbana-
Champaign, Urbana, IL 61801 USA (e-mail: varshney@illinois.edu).  He is also with Salesforce Research, Palo Alto, CA, USA.}}
\author{Avhishek Chatterjee and Lav R. Varshney, \IEEEmembership{Senior Member, IEEE}
}

\begin{document}

\maketitle

\begin{abstract}
Due to energy-efficiency requirements, computational systems are now being implemented using noisy nanoscale semiconductor devices whose reliability depends on energy consumed.  We study circuit-level energy-reliability limits for  deep feedforward neural networks (multilayer perceptrons) built using such devices, and en route also establish the same limits for formulas (boolean tree-structured circuits).  To obtain energy lower bounds, we extend Pippenger's mutual information propagation technique for characterizing the complexity of noisy circuits, since small circuit complexity need not imply low energy. Many device technologies require all gates to have the same electrical operating point; in circuits of such uniform gates, we show that the minimum energy required to achieve any non-trivial reliability scales superlinearly with the number of inputs. Circuits implemented in emerging device technologies like spin electronics can, however, have gates operate at different electrical points; in circuits of such heterogeneous gates, we show energy scaling can be linear in the number of inputs. Building on our extended mutual information propagation technique and using crucial insights from convex optimization theory, we develop
an algorithm to compute energy lower bounds for any given boolean tree under heterogeneous gates. This algorithm runs in linear time in number of gates, and is therefore practical for modern circuit design.  As part of our development we find a simple procedure for energy allocation across circuit gates with different operating points and neural networks with differently-operating layers. 
\end{abstract}

\section{Introduction}
\label{sec:intro}
As neural networks become larger and more prevalent, their energy requirements are becoming of key concern \cite{SchwartzDSE2019_arXiv}.  Though most current deep networks are enormous cloud-based structures, there is a further desire for hardware implementations for mobile, in-sensor, and in-memory inference \cite{KangKSEC2014,WangLV2015,ZhangS2015,AhnHYMC2015}.  At the same time, as area and energy scaling in CMOS technology saturates, the semiconductor industry has been exploring promising new energy-efficient nanoscale devices as computational substrates \cite{ShanbhagVKPV2019}.
Special-purpose nanoscale hardware is
faster and more energy-efficient than alternate approaches, making deep learning suitable for applications ranging from voice recognition on mobile devices to in-sensor health monitoring \cite{Neftci2016,Shanbhag2016_arXiv}. 
A major challenge, however, in using nanoscale devices is that they can be very unreliable, especially when operated at low energy \cite{ButlerMMVRRH2012}. This has renewed interest in the study of reliable circuit design using unreliable components, both digital and analog \cite{ShanbhagMVOMRJR2008,GuptaADDGKMNRSSS2013,NahlusKSB2014,De2016}, a problem first addressed by von Neumann through a modular redundancy approach \cite{VonNeumann1956}. 

For the success of low-power inference, understanding energy-reliability limits of nanoscale neural networks is important.  
Nanoscale devices fail at random, but for each particular device technology like spintronics or carbon nanotubes, there is a functional relationship between failure probability and energy.
Devices consume more energy as they are built to have lower failure rates \cite{Kish2004,ButlerMMVRRH2012}.  Here we aim to use the device-level relationship to determine basic energy-reliability limits at the circuit level. We focus on deep feedforward networks (multilayer perceptrons), which are directed acyclic graphs (DAGs).  As a simpler setting to build towards neural networks, we also consider tree-structured boolean circuits (often known as \emph{formulas}), which are of independent interest in fault-tolerant computing.

Past information-theoretic studies have focused on bounding the minimum size of a noisy circuit to compute a function with a given reliability, e.g.~\cite{DobrushinO1977,PippengerST1991,GacsG1994}, and to upper bound the device noise for which a non-trivial reliability can be achieved \cite{HajekW1991,Pippenger1988,EvansS1998}, largely restricted to formulas.  Using a basic mathematical tool due to Pippenger \cite{Pippenger1988}, the lower bound on circuit size for a target reliability has been improved \cite{Evans1994,HajekW1991, EvansS1999}.  Such circuit complexity results, however, do not directly provide insight into the basic {\em energy} requirements for reliable nanoscale circuits as we aim to obtain here.  After all, one could consider making a fixed number of individual gates less noisy with more energy, or one could construct larger and more redundant circuit designs with gates that remain noisy.  Determining best design strategies is useful not just as a proof technique, but also for informing practical circuit design and explaining the nature of biological neural networks in sensory cortex, as we detail in separate works \cite{ShanbhagVKPV2019, ChatterjeeV2019}.  In neurobiology, we show that neural connectivity, reliability,  and energy characteristics are matched to one another, as per our theory \cite{ChatterjeeV2019}. In circuit design, \cite{ShanbhagVKPV2019} details several circuits that were designed for practical problems.

Here we extend Pippenger's mutual information propagation technique and use crucial insights from convex optimization theory to determine energy limits for reliable nanoscale boolean trees and feedforward neural networks with unreliable components.  The main contributions are detailed next. 

\subsection{Contributions}
En route to neural network results, we first derive energy-reliability limits for boolean trees where all logic gates are constrained to have uniform electrical operating points (Sec.~\ref{sec:uniform}), a constraint common in many extant device technologies. The main goal is to understand the scaling of energy consumption with number of inputs. We observe that a superlinear scaling of energy consumption with number of inputs is unavoidable for both extant and emerging technologies.

As some new technologies promise to relax the constraint of uniform operating points, we also study heterogeneous logic gates that consume different energies (Sec.~\ref{sec:nonuniform}). Extending the mutual information propagation technique and using ideas from convex optimization theory, we determine the minimum energy needed for a given reliability requirement and vice versa. For certain symmetric circuits, linear scaling of energy consumption is possible. We also obtain an efficient procedure for energy allocation.

Note that in the presence of unreliable components, minimum complexity realizations may not consume the minimum energy. Hence, we must go beyond
previous work on circuit complexity bounds in fault-tolerant computing. Unlike previous work 
which gives complexity bounds only on the class of all $n$-input circuits, we also
propose a linear-time algorithm to bound energy for any given boolean 
tree under non-uniform gate operations. This algorithm draws on crucial 
insights from convex optimization theory 
and is specifically useful in practice for modern circuits with very large
numbers of gates, see \cite{ShanbhagVKPV2019}. This method also leads to a heuristic energy allocation scheme. This is because the optimization problem corresponding to the bound can be seen as a convex surrogate for the exact energy allocation problem, which is intractable in general.

Returning to feedforward neural networks, we study two complementary scenarios: (i) all neurons in the network have 
{\em uniform} energy consumption, and (ii) neurons in the same layer
have the same energy consumption, but {\em non-uniform} energy consumption across layers. 
We again build on the mutual information propagation technique (now extended to consider DAGs) to obtain energy bounds that yield insights into the structural and connectivity
requirements for reliable operations of nanoscale feedforward neural networks. We also obtain a design
heuristic for choosing operating points in a deep neural network---a simple energy allocation that informs practical circuit design.

Note that earlier presentations of this work \cite{ChatterjeeV2016,ChatterjeeV2017a} were focused only on circuits with homogeneous electrical operating points for devices, whereas the new synthesis in the current paper emphasizes the value of heterogeneous operation of gates and neural network layers. Results on energy allocation are therefore novel to this paper.

\subsection{Related Work}

A recent paper on energy-efficient circuit design \cite{AntoniadisBNPS2014}  is similar
to our work in obtaining energy bounds for reliable computing, but there are notable differences. 
First, in terms of mathematical approach, we extend a mutual information propagation technique \cite{Pippenger1985}, whereas they build on a circuit equivalence technique \cite{DobrushinO1977,GacsG1994}. 
Second, our approach to obtaining energy bounds for a given formula circuit offers design insights on good energy allocation in that circuit.  Such quantitative design insights are very useful to circuit design practitioners, e.g.\ \cite{ShanbhagVKPV2019}. 
 Third, our approach works for any convex smooth energy-failure function
for gates, whereas \cite{AntoniadisBNPS2014} requires strong assumptions on the
energy-failure relations. Our bounding techniques and design insights can also be 
extended to the case where devices in a circuit have different energy-failure functions.
Finally, most importantly, we can consider directed acyclic computation graphs in the context of feedforward neural networks, which move beyond just tree-structured formula circuits.

There are some further differences: our approach yields energy bounds for any particular formula circuit (rather than just bounds over the class of all $n$-input formula circuits),  we develop a linear-time (in $n$) algorithm  that computes the bound, and we demonstrate effectiveness of our design heuristic on simple circuits.

\section{Models}
This section provides mathematical models for formulas and for feedforward neural networks.

\subsection{Boolean Formulas}
\label{sec:model}
The goal is to design circuits to compute $n$-input boolean functions using a single type of gate among the set of universal gates, i.e.\ \textsc{nand} and \textsc{nor}, such that each gate has at most $k$ inputs and exactly one output. Gates are interconnected into a circuit to compute the desired function, such that inputs to gates are some of the $n$ inputs to the function, outputs of other gates in the circuit, or constants $\{0,1\}$.  The output of the circuit is the output of a certain gate. In general, a boolean function can be realized by several different circuits using the same kind of universal gate. Indeed, elementary digital logic design is concerned with minimal realizations \cite{Karnaugh1953}.

We assume that the boolean function $F$ is sensitive to each input, i.e., for each input $i$, there is a configuration of other inputs $x_1=c_1, x_2=c_2, \ldots, x_{i-1}=c_{i-1}, x_{i+1}=c_{i+1}, \ldots, x_n=c_n$, such that
\[
F(c_1, \ldots, c_{i-1}, 0, c_{i+1}, c_n) \neq F(c_1, \ldots, c_{i-1}, 1, c_{i+1}, c_n)\mbox{.}
\]
A boolean function of $n$ inputs which is not sensitive to one of its inputs is equivalent to a function of $n-1$ inputs.
In  combinatorial circuits there is no feedback, i.e., the connections between the gates must form a directed acyclic graph, where gates are vertices and the connections between gates are edges. The input to a gate is considered the head of a directed edge.  A special class of combinatorial circuits of interest are \emph{formulas}, where the graph is a directed tree.

Under the $\epsilon$-noisy model \cite{VonNeumann1956},
a gate produces a correct output for a given input with probability $1 - \epsilon$ and flips its output with probability $\epsilon$. Each gate in a circuit fails independently of any other. 
The noise probability $\epsilon$ of a gate depends on its electrical characteristics such as bias voltage, as well as physical and material properties. These characteristics also determine the energy consumption of the gate.  For a given device technology, there is a relationship between energy consumption of the gate and its probability of failure that depends on the fundamental nature of the device, whether CMOS or emerging beyond-CMOS technologies such as spin electronics or carbon nanotubes.
\begin{definition}
For a gate with probability of failure $\epsilon$ and energy consumption $e_g$, let the \emph{energy-failure function} of the gate be $\epsilon=\chi(e_g)$. 
\end{definition}

When we construct circuits from noisy gates, for any given input, there is a probability that the output of a circuit is incorrect. 
\begin{definition}
We say a circuit for the $n$-input boolean function $F$ is \emph{$\delta$-reliable} if for any input configuration $\{x_1, x_2, \ldots, x_n\} \in \{0,1\}^n$, the output of the circuit $y$ satisfies the following: 
$$\Pr(y=F(x_1,x_2,\ldots,x_n))\ge 1-\delta\mbox{,}$$
where probability is over all the failure patterns of the gates in the circuit.
\end{definition}

This work aims to answer the following general question: What is the minimum energy needed to realize an $n$-input boolean function using a $\delta$-reliable formula? We first consider circuits (and/or technologies) where each gate in the circuit must have the same operating point and hence consume the same energy.  We obtain the minimum energy per device for reliable computation. Some emerging technologies like spin electronics allow different devices in the circuit to have different operating points \cite{ShanbhagVKPV2019}.  The energy-reliability limit with such heterogeneous gates is better; we characterize such settings later.

\subsection{Feedforward Neural Networks}
\label{sec:nndef}
Consider a binary $L$-layer feedforward neural network with $n$ inputs and a single output
used to learn and approximate potentially complicated logic functions---a so-called multilayer perceptron---as described in 
standard textbooks \cite{RussellN2010}. 
Inputs to the neural network are $\pm 1$, where logical $0$ maps to $-1$ and logical $1$ maps to $+1$.
The neural network has a given connection pattern between neurons in the various layers, e.g.\ full connectivity or
$d$-regular connectivity between layers, and there are real-valued weights on the edges.
Each neuron has an associated activation function with a real-valued input and an output
that is $\pm 1$. For a neuron, the input to the activation function is a sum of the outputs of the neurons
of the previous layer, weighted by the edge weights. 

A neuron $g$ can fail independently of any other neuron with probability $\epsilon_g$. When
a neuron fails, it flips the output from $+1$ to $-1$ and vice versa. 
As in boolean circuits, neuron energy consumption
$e_g$ and $\epsilon_g$ are related by energy-failure function $\chi$. 
We define the notion of a $\delta$-reliable neural network in the same
way and ask: What is the minimum energy expenditure in a $\delta$-reliable
neural network?

\section{Preliminaries}
\label{sec:preliminaries}
We present some preliminary definitions and results for the boolean tree problem.

\subsection{Circuit Graphs}
An $n$-input boolean function $F_n$ can be realized by various different formula structures of (universal) logic gates of a given kind. For a given realization, we have a directed graph $G_g=(V_g,\mc{E}_g)$, where vertices $V_g$ are the set of gates and $\mc{E}_g$ are the edges corresponding to connections between gates. Each edge is directed towards the gate to which it is an input. We call $G_g$ the \emph{gate graph} of the realization of the formula. In this section we only consider formulas, and hence, the corresponding graphs are trees.

Another graph that we use subsequently is the \emph{bit graph} $G_b=(V_b,\mc{E}_b)$. Here, $V_b$ corresponds to interconnects/wires in the circuit. In a circuit there are interconnects that run from output of one gate to input of another gate, from circuit inputs to inputs of gates, and from fixed sources (corresponding to permanent $0$ or $1$) to input of gates. Two nodes in $V_b$ share an edge if the corresponding interconnects are incident to a common gate. As the circuit is a tree, interconnects between gates have a one-to-one mapping to gates (the gates to which they are outputs) and further $G_g$ is a subgraph of $G_b$. In addition $G_b$ has leaf nodes that correspond to either inputs or to fixed sources. This relation will be useful later.

\subsection{Mutual Information Propagation}
Pippenger developed the technique of circuit information flow to bound the size of a $\delta$-reliable circuit constructed using $\epsilon$-noisy gates \cite{Pippenger1988}, which we review.  For a boolean function $F$ of $n$ inputs, for any input $i$ there exist values $c_{\backslash i} \in \{0,1\}^{n-1}$ such that $F_i(x):=F(c_{\backslash i}, x) = x$ or $\bar{x}$, where $\bar{x}$ is the complement of $x$. Hence, for any random input $X$, $F_i(X)$ is a one-to-one mapping. When $F$ is realized using noisy gates, the output is no longer $F_i(X)$, but a random variable $Y$, depending on $G_g$ and gate noise. 

Note $\delta$ is the upper bound on probability of error, $P_e(c_{\backslash i})$ for input configuration $c_{\backslash i}$, and the possible number of values of $Y$ is $M=2$. Then using Fano's inequality:
\begin{equation}
I(X;Y) \ge H(X) - h(P_e(c_{\backslash i})) - P_e(c_{\backslash i}) \frac{\ln(M-1)}{\ln 2} \mbox{,}
\end{equation}
where $h(\cdot)$ is the binary entropy function.  Since $M=2$ as the output is binary, it follows that:
\begin{equation}
1-h(\delta) \le I(X;Y)\mbox{.}
\end{equation}
Unlike in communications, we do not bound $h(P_e)$ by $1$ to simplify.  

Finally to bring gate failures into this bound, two observations are made. First, for a perfect gate $g$ with input random variables $U_1, U_2, \ldots, U_k$ and output random variable $U_0$, for any random variable $Z$:
\begin{align}
I(Z;U_0) & \le I(Z;U_1, U_2, \ldots, U_k) \nonumber \\
& \le \sum_{i=1}^k I(Z;U_i) \label{eq:gateCombine}
\end{align}
by the data processing inequality and the distributive rule of mutual information \cite{Pippenger1988}.
Second, if the gate is $\epsilon$-noisy, then the output is $\tilde{U}_0=U_0+N_0$, where $N_0$ is a Bernoulli($\epsilon$) random variable independent of $U_0$. Then, a strong data processing inequality holds \cite{Pippenger1988,Evans1994,EvansS1999}:
\begin{align}
I(Z;\tilde{U}_0) \le (1-2\epsilon)^2 I(Z;U_0).\label{eq:gateError}
\end{align}

Based on these two information inequalities and some combinatorial arguments, we get a lower bound on the depth of a formula in terms of $\delta$ and $\epsilon$ \cite{Pippenger1988,Evans1994,EvansS1999}. There,  gate noise was fixed and energy consumption was not considered. Here, we build on this technique  to study circuits where energy (hence, noise) of a gate can be tuned and derive bounds on total energy consumption to realize a boolean function.

\subsection{Energy-Failure Functions}
A few special cases of the energy-failure function $\chi$ are relevant to modern device technologies. Here we will obtain a generic lower bound applicable to a broad class of energy-failure functions that encompass all relevant technologies.  We define a class of functions as follows.
\begin{definition}
Let  \emph{physical} energy-failure functions be ones that satisfy $\chi:(0,\infty) \mapsto (0,a)$, $0<a\le 1$, that is strictly decreasing, convex, and differentiable with $\lim_{e_g\to 0} \chi(e_g)=a$ and $\lim_{e_g \to \infty} \chi(e_g) = 0$. 
\end{definition}
\begin{lemma}
\label{lem:inverseChi}
Any physical energy-failure function $\chi$ has an inverse $\chi^{-1}$ that is strictly decreasing, convex, and differentiable.
\end{lemma}
\begin{IEEEproof}
See Appendix. 
\end{IEEEproof}

Energy-failure functions for CMOS, carbon nanotube, and spin electronics are all physical.
A closed-form expression relating energy and failure for a typical spin device has been derived based on the physics of the device \cite{ButlerMMVRRH2012}. An approximate form of the functional dependence is $\epsilon=\epsilon_0 \exp(-c I)$,
where $I$ is the supply current of the device. Constants $c>0$ and $\epsilon_0 \in (0,1]$ depend on the device parameters, like critical current, gate delay, and other switching parameters. As energy consumption scales as square of current, failure $\epsilon$ and energy $e$ in a spin device are related as: $\epsilon=\epsilon_0 \exp(-c' \sqrt{e})$, with $c'>0$.
A generic way to capture this kind of dependence is through stretched exponentials, $\epsilon = \epsilon_0 \exp(-c e^\beta)$,
where $\epsilon_0, \beta \in (0,1]$, and $c>0$. 
For CMOS technologies, it has been shown that the exponential energy-failure function is a fundamental thermodynamic limit \cite{Kish2004}.
Polynomial functions are another wide class of energy-failure dependence functions that can be used to approximately characterize different logic devices: $\epsilon = \frac{\epsilon_0}{(e+1)^\beta}\mbox{, with } \beta>0, \epsilon_0 \in (0,1)$.

We now proceed to determine the energy-reliability limits of boolean trees and will then return to 
energy-reliability limits of neural networks.

\section{Boolean Tree Circuits from Homogenous Gates}
\label{sec:uniform}
Let us consider the setting where each gate in the circuit must have the same electrical operating point. This may be due to limitations of the design and fabrication technologies of the electronic devices being used. Some emerging technologies do allow variable gate operations \cite{ShanbhagVKPV2019}, which we discuss in Sec.~\ref{sec:nonuniform}. 

\subsection{Computation Energy per Input Bit}
For an $n$-input formula $F$, consider a given realization of the formula using logic gates and hence a given directed gate tree $G_g=(V_g,\mc{E}_g)$.  For each input bit $i$, 
$
1-h(\delta) \le I(F_i(X);X) \le (1-2\epsilon)^{2|\mc{P}_i|}$,
where $\mc{P}_i$ is the path in $G_g$ to the output gate from the gate to which $x_i$ is input.   This follows by inductively using  \eqref{eq:gateCombine} and \eqref{eq:gateError} along the depth of $G_g$ from root to the terminal gate into which $x_i$ is an input \cite{Pippenger1988,Evans1994,EvansS1999}.

Thus, for any $\delta$-reliable circuit realization of $F$ with total energy consumption $E$ the following conditions must be satisfied.
\begin{align}
\mbox{C1.} &\quad 1-h(\delta) \le (1-2\epsilon)^{2|\mc{P}_i|}\mbox{, for } 1 \le i \le n, \label{eq:uniformCond1}\\
\mbox{C2.} &\quad \chi\left(\frac{E}{|V_g|}\right)=\epsilon. \nonumber 
\end{align}
Condition C1 follows from the definition of $\delta$-reliability of the circuit since for any input bit $i$ and any configuration $c_{\backslash i}$, $\Pr(Y=F_i(X)) \ge 1-\delta$. Condition C2 relating total energy $E$ and $\epsilon$ follows because gates with the same electrical characteristics consume the same energy. Hence, energy consumption per gate is $e_g=\frac{E}{|V_g|}$, which along with $\epsilon=\chi(e_g)$ implies the condition.

Our goal is to find a lower bound on the total energy consumption $E$ in a circuit realization of $F$. This implies that a condition weaker than C1, together with C2, would give a lower bound on $E$. Hence, we develop the following weaker condition.
\begin{align}
\mbox{C3.} &\quad \frac{1}{4} \ln\frac{1}{1-h(\delta)} \ge |\mc{P}_i| \epsilon\mbox{, for } 1 \le i \le n, \label{eq:uniformCond2}
\end{align}

\begin{lemma}
\label{lem:weakerCond}
For a given formula $F$, realization gate tree $G_g=(V_g,\mc{E}_g)$, and required reliability $\delta$, if $E$ satisfies conditions (C1, C2), then it also satisfies conditions (C3, C2). 
\end{lemma}
\begin{IEEEproof}
To relate C1 and C3, we note that 
$1-x \le \exp(-x)$,
which implies that if \eqref{eq:uniformCond1} is satisfied for a given $E$, then for that $E$:
$
1-h(\delta)\le \exp(-4|\mc{P}_i| \epsilon)$, for $ 1\le i\le n
$.
Taking the natural logarithm of both sides does not disturb the inequality since $\ln(\cdot)$ is monotonic. Thus we obtain the desired result by multiplying both sides by $-1$ and flipping the inequality. 
\end{IEEEproof}
Due to linear dependence on graph structure, condition C3 is more tractable than C1. 
As we discuss later, these conditions lead to a closed-form lower bound on energy consumption.

It is apparent from these necessary conditions that the lower bound on energy will depend on the $\chi$ function. Intuitively, when probability of failure of a gate decays faster with its energy consumption, the total energy consumption of the circuit should also be lower. 

\begin{lemma}
\label{lem:eLBuniformGraph}
For physical energy-failure functions $\chi$ and for a given $n$-input boolean function $F$ with realization graph $G_g=(V_g,\mc{E}_g)$,
\begin{align}
E \ge |V_g| \chi^{-1}\left(\frac{1}{4 \max_i |\mc{P}_i|} \ln\frac{1}{1-h(\delta)}\right) \mbox{, for } \delta<\frac{1}{2}. \nonumber
\end{align}

\end{lemma}
\begin{IEEEproof}
For a given $n$-input formula $F$ and its corresponding directed tree $G_g=(V_g,\mc{E}_g)$, it follows from \eqref{eq:uniformCond2} that
\begin{align}
\epsilon \le \min_i \frac{1}{4 |\mc{P}_i|} \ln\frac{1}{1-h(\delta)}  = \frac{1}{4 \max_i |\mc{P}_i|} \ln\frac{1}{1-h(\delta)}. \nonumber
\end{align}

Now as $\epsilon = \chi(E/|V_g|)$, and $\chi^{-1}$ is strictly decreasing, $\epsilon  \le  \frac{1}{4 \max_i |\mc{P}_i|} \ln\frac{1}{1-h(\delta)}$ implies that for a given formula and $G_g$,
\begin{align}
E \ge |V_g| \chi^{-1}\left(\frac{1}{4 \max_i |\mc{P}_i|} \ln\frac{1}{1-h(\delta)}\right). \nonumber
\end{align}
\end{IEEEproof}

We also state some combinatorial properties of trees with a fixed number of leaves, proved in the Appendix, that allow us to
give the main theorem on minimum energy per input bit.
\begin{lemma}
\label{lem:leafSize}
Among all directed rooted trees with $L$ leaves and number of children bounded by $k$, a $k$-ary balanced tree has the minimum number of non-leaf nodes.
\end{lemma}
\begin{lemma}
\label{lem:leafDepth}
Among all directed rooted trees with $L$ leaves and number of children bounded by $k$, a $k$-ary balanced tree has the minimum depth for the subtree made of non-leaf nodes.
\end{lemma}
\begin{lemma}
\label{lem:leafMonotone}
Among the class of directed rooted trees with bounded children, the minimum depth and minimum size of the subtree consisting of non-leaf nodes are monotone in number of leaves.
\end{lemma}

\begin{theorem}
\label{thm:eLBuniform}
The minimum energy required to realize any $n$-input boolean function using a $\delta$-reliable ($\delta < \frac{1}{2}$) formula of gates with degree  no more than $k$ (with $k<n$), and each with the common physical energy-failure function $\chi$ is 
\[
\frac{n}{k} \chi^{-1}\left(\frac{\ln k}{4\ln n} \ln\frac{1}{1-h(\delta)}\right)\mbox{.}
\]
\end{theorem}
\begin{IEEEproof}
A lower bound on energy consumption over all realization circuits (graphs) and all $n$-input boolean functions can be obtained by minimizing the realization-specific bound from Lemma \ref{lem:eLBuniformGraph} over all realizations and $n$-input functions: 
\begin{align}
E &\ge \min_{F_n, G_g} |V_g| \chi^{-1}\left(\frac{1}{4 \max_i |\mc{P}_i|} \ln\frac{1}{1-h(\delta)}\right) \nonumber \\
&\ge \left(\min_{F_n, G_g} |V_g|\right) \min_{F_n, G_g} \chi^{-1}\left(\frac{1}{4 \max_i |\mc{P}_i|} \ln\frac{1}{1-h(\delta)}\right) \nonumber \\
&= \left(\min_{G_g: n\text{ inputs}} |V_g|\right) \min_{G_g: n\text{ inputs}} \chi^{-1}\left(\frac{1}{4 \max_i |\mc{P}_i|} \ln\frac{1}{1-h(\delta)}\right), \nonumber
\end{align}
where the last equality follows because for each formula there is a gate tree and vice versa.

For a rooted tree $G=(V,\mc{E})$, let $\ell(V)$ and $\bar{\ell}(V)$ denote the leaf and non-leaf nodes, respectively. Then, it is clear from the relationship between bit graphs and gate graphs that $\bar{\ell}(V_b)=V_g$ and $\mc{E}_g=\mc{E}_b \cap (V_g \times V_g)$.

As $G_g$ has $n$ inputs, the corresponding $G_b$ has at least $n$ leaves. Thus, one can write
\begin{align}
& \min_{G_g: n\text{ inputs}} |V_g| = \min_{G_b: \ge n \text{ leaves}} \bar{\ell}(V_b). \nonumber
\end{align}
By Lemma~\ref{lem:leafMonotone} and the fact that adding more constraints only increases the minimum we have
$$\min_{G_b: \ge n \text{ leaves}} \bar{\ell}(V_b) \le 
\min_{G_b: n \text{ leaves}} \bar{\ell}(V_b).$$

By Lemma~\ref{lem:leafSize}, a $k$-ary $G_b$ achieves the minimum. Now as there are $n$ leaves, in a $k$-ary tree there are at most $\lceil \frac{1}{k} n \rceil$ nodes in the level above. In turn, there are at most
$\lceil \frac{1}{k}\lceil \frac{1}{k} n \rceil \rceil$ in the level above that and so on. This continues until we have only one node at the top level. Thus the total number of non-leaf nodes are lower bounded by 
$
\lceil \frac{1}{k} n \rceil + \lceil \frac{1}{k}\lceil \frac{1}{k} n \rceil \rceil + \cdots + 1 \ge \frac{n}{k}
$.
This implies that $\min_{F, G_g: n \text{ inputs}} |V_g| \ge 
\frac{n}{k}$.

To bound the other term note that $\chi^{-1}$ is strictly decreasing, so  $\chi^{-1}\left(\frac{1}{4 \max_i |\mc{P}_i|} \ln\frac{1}{1-h(\delta)}\right)$ is minimized when $\frac{1}{4 \max_i |\mc{P}_i|} \ln\frac{1}{1-h(\delta)}$ is maximized. This is because $\ln\frac{1}{1-h(\delta)}\ge 0$, for $\delta<\frac{1}{2}$. Thus when $\max_i |\mc{P}_i|$ is minimized the other term is also minimized.

As $\max_i |\mc{P}_i|$ is the depth of $G_g$, by Lemma \ref{lem:leafDepth}, this minimum is achieved by a $k$-ary tree. Now, by Lemma \ref{lem:leafMonotone} and the relation between $G_b$ and $G_g$, the depth of $G_g$ is minimized when the number of leaves in $G_b$ is minimized. The number of leaves in $G_b$ is no less than $n$, as there are $n$ inputs. For a $k$-ary tree with $n$ leaves, the depth is at least $\lceil \frac{\ln n}{\ln k} \rceil \ge \frac{\ln n}{\ln k}$. 
Hence,
\[
E \ge \frac{n}{k} \chi^{-1}\left(\ln\frac{1}{1-h(\delta)}\frac{\ln k}{4\ln n}\right)\mbox{.}
\]
\end{IEEEproof}

To understand the implications of Theorem \ref{thm:eLBuniform}, let us consider the simple case where we fix $\delta \in (0,\frac{1}{2})$ and $k=O(1)$. Then, the theorem implies that to realize $\delta$-reliability using gates of at most $k$ inputs, the minimum energy requirement scales with number of inputs $n$ as
\[
\Omega\left(n \chi^{-1}\left(\frac{c(\delta,k)}{\ln n}\right)\right)\mbox{, for some constant } c(\delta,k).
\]

Typical nanoscale circuits compute functions of large numbers $n$ of inputs and so such order scaling is of central interest.
In most practical scenarios (including the conditions in Theorem \ref{thm:eLBuniform}), $\chi^{-1}$ is strictly decreasing and $\lim_{\epsilon \to 0} \chi^{-1} = \infty$. Thus, as $n \to \infty$, the term
$
\chi^{-1}\left(\frac{c(\delta,k)}{\ln n}\right) \to \infty
$. 
This implies that the minimum energy requirement per input bit over all boolean functions increases with number of input bits for any $\delta \in (0,\frac{1}{2})$ and $k=O(1)$.
  
This contrasts sharply with circuits of noiseless gates, where there are many $n$-input functions that can be realized using $O(n)$ gates. As each gate requires only a constant amount of energy for its perfect operation, the total energy consumption is $O(n)$ for these circuits. Hence, as devices and gates become unreliable there is a price to be paid in terms of the energy per input bit.

Next, we build a more quantitative understanding of the scaling of energy per input bit by considering a few relevant classes of energy-failure dependence functions $\chi$.

\subsection{Energy Bounds for Device Technologies}
\label{sec:spin}
As discussed before, the energy-failure functions for most device technologies are either
polynomial or stretched exponential.
The following lower bounds for polynomial and stretched exponential energy-failure dependence follow from Theorem \ref{thm:eLBuniform}.
\begin{corollary}
\label{cor:polyExp}
The minimum energy required to realize any $n$-input boolean function using gates with maximum $k$ inputs, $k <n$, and each with $\epsilon=\chi(e)$ is 
\[
\frac{n}{k c^{\frac{1}{\beta}}} \left(\ln\frac{4\epsilon_0\ln n}{\ln k} - \ln\ln\frac{1}{1-h(\delta)}\right)^{\frac{1}{\beta}}\]
when $\chi(e)=\epsilon_0 \exp(-c e^\beta)$, and
\[
\frac{n}{k} \left(\frac{4 \epsilon_0 \ln n}{\ln k \ln\frac{1}{1-h(\delta)}}\right)^{\frac{1}{\beta}}
\]
when $\chi(e)=\frac{\epsilon_0}{(e+1)^\beta}$.
\end{corollary}
\begin{IEEEproof}
Since both energy-failure functions are physical, the result follows by substituting the appropriate $\chi^{-1}$ in Theorem \ref{thm:eLBuniform}.
\end{IEEEproof}

When we have a lower bound on $\chi$ rather than an exact functional form, we can still obtain a lower bound on energy consumption. 
\begin{lemma}
\label{lem:ChiEMonotone}
Let energy-failure functions $\chi_1$ and $\chi_2$ be physical and $\chi_1(e) \le \chi_2(e)$ for all $e$. For gates with $\chi_1$ if there exists no $\delta$-reliable circuit $G_g$ for a formula $F$ with total energy consumption no more than $E$, then the same is true for $\chi_2$ gates.
\end{lemma}
\begin{IEEEproof}
The result follows by noting that if a per-gate energy $e_2$ in case of $\chi_2$ achieves $\delta$-reliability, then by the monotonicity of $\chi_1$ and $\chi_2$, the dominance between them, and condition \eqref{eq:uniformCond2}, $e_2$ energy per gate achieves $\delta$-reliability in case of $\chi_1$ gates.
\end{IEEEproof}
Thus, if we have a lower bound on $\chi$, we can find the fundamental lower bound on energy consumption. This is useful when device physics are not tractable and an exact functional form is unknown. Here we use this property to make an interesting generic observation about a broad class of $\chi$ functions.

Note that the exponential function $\epsilon = \epsilon_0 \exp(-c e)$ lower bounds polynomial and stretched exponential functions. Thus, by Lemma \ref{lem:ChiEMonotone}, the exponential function can be used to obtain a lower bound on energy consumption for all sub-exponential energy-failure functions. Since energy-failure functions for many devices fall in the sub-exponential class and the exponential energy-failure function is also a fundamental thermodynamic limit for CMOS technologies, we can obtain a generic bound using exponential functions. 

From Corollary \ref{cor:polyExp} it follows that for $\epsilon=\epsilon_0 \exp(-c e)$, the energy consumption for computing $n$-input functions is lower bounded by
\begin{align}
&  \frac{n}{c k} \left(\ln\ln n - \ln\ln k + \ln (4\epsilon_0) - \ln\ln\frac{1}{1-h(\delta)}\right). \label{eq:loglogn}
\end{align}

As $k=O(1)$ and so is $\delta$, for large $n$ the leading term
is $\frac{n \ln\ln n}{c k}$. So, for large $n$, as long as $\delta<\frac{1}{2}$, the minimum energy requirement scales at least as $\frac{n \ln\ln n}{c k}$. This has important implications.

First, if gates are error-prone and have sub-exponential energy-failure functions, then to achieve any non-trivial reliability ($\delta<\frac{1}{2}$), the energy requirement per bit of computation scales at least as $c' \ln\ln n$, and this lower bound on scaling is independent of the reliability requirement (as long as it is non-trivial). This means that the reliability requirement is not the bottleneck in obtaining linear scaling of energy with number of inputs. Rather, the bottleneck is the sub-exponential unreliability of the gates.

Second, even if we allow a decreasing reliability requirement with increasing input size, it does not help in obtaining a constant energy consumption per input bit. As long as $\delta_n \uparrow \frac{1}{2}$ such that
$h(\delta_n) = 1 - \omega(\frac{1}{n})$, the minimum energy requirement per input bit scales at least as $\ln\ln n$, irrespective of the scaling of $\delta_n$. This can be seen by substituting $h(\delta_n)$ for $h(\delta)$ in \eqref{eq:loglogn}. This further implies that the fundamental bottleneck is the sub-exponential unreliability of the gates, not the reliability requirement.

\section{Boolean Tree Circuits from Heterogeneous Gates}
\label{sec:nonuniform}
Emerging technologies like spin electronics promise circuits with gates having different electrical operating points (and energy levels). 
Now we determine the minimum energy required to compute a given $n$-input boolean function using a given gate graph if the energy of each gate is tuned separately. 

Let $F$ be an $n$-input boolean function with a gate (realization) graph $G_g$, a directed rooted tree. We allocate energy to each gate to ensure that the circuit is $\delta$-reliable and is also energy efficient. Our goal is twofold: to characterize the minimum energy required by a circuit realization while having $\delta$-reliability and to understand the best reliability that can be achieved for a given energy budget.

Using information propagation we find that to achieve $\delta$-reliability,
for each input bit $i$, we need: 
\begin{equation}
1-h(\delta) \le I(F_i(X);X) \le \prod_{g \in \mc{P}_i}(1-2\epsilon_g)^{2}\mbox{,} \nonumber
\end{equation}
where $\mc{P}_i$ is the path in $G_g$ to the output gate from the gate to which $x_i$ is input.  As in the case of uniform gates, this follows by inductively using  \eqref{eq:gateCombine} and \eqref{eq:gateError} along the depth of $G_g$ from root to the terminal gate into which $x_i$ is an input. After straightforward algebraic manipulations, the condition becomes
\begin{equation*}
\frac{1}{4} \ln\frac{1}{1-h(\delta)} \ge \sum_{g \in \mc{P}_i} \epsilon_g,
\end{equation*}
where $\epsilon_g = \chi(e_g)$ for each $g \in \mc{V}_g$. We introduce the notation
$\gamma(\delta)=\frac{1}{4} \ln\frac{1}{1-h(\delta)}$. Note that as $\epsilon_g \ge 0$, to satisfy the necessary condition for $\delta$-reliability, we only need to consider the {\em maximal paths} $\{\mc{P}_i\}$, where a path $\mc{P}_i$ for an input bit $i$ is maximal if there exists no bit $j\neq i$ such that $\mc{P}_i \subset \mc{P}_j$.
Hence, the necessary condition for $\delta$-reliability is 
\begin{equation}
\gamma(\delta) \ge \sum_{g \in \mc{P}} \epsilon_g, \mbox{ for all maximal }
\mc{P}. \label{eq:neceNonUni}
\end{equation}
In a tree $G_g$, the only maximal paths are the unique {\em root-to-leaf} paths for each leaf node in $G_g$. 

\subsection{Minimum Energy Requirement}
\label{sec:minEnergyNonUni}
Based on the necessary condition \eqref{eq:neceNonUni}, we first determine
the minimum energy requirement for  $\delta$-reliable realization of $F$ using 
$G_g$. Consider the following problem,
\begin{align}
& \min \sum_{g \in V_g} e_g \nonumber \\
& \mbox{s.t.}~\sum_{g \in \mc{P}} \epsilon_g \le \gamma(\delta) \mbox{ for all maximal } \mc{P}, \nonumber \\
& \epsilon_g = \chi(e_g)\mbox{ and }e_g\ge 0~\forall g \in V_g. \label{eq:minEnonUni1}
\end{align}

The constraints in this optimization problem are from the $\delta$-reliability requirements and
the objective is the total energy consumed across gates.
Thus, the optimal solution satisfies the necessary conditions
for $\delta$-reliability and is a lower bound on the total energy required to realize $F$ using $G_g$. 

For physical energy-failure functions, $\chi$ has an inverse, $0\le \epsilon_g=\chi(e_g)\le a$, for $0\le e_g \le \infty$.
Hence, the  problem can be further simplified by eliminating the energy 
variable $e_g$,
\begin{align}
& \min \sum_{g \in V_g} \psi(\epsilon_g) \nonumber \\
& \mbox{s.t.}~\sum_{g \in \mc{P}} \epsilon_g \le \gamma(\delta) \mbox{ for all maximal } \mc{P}, 
\nonumber \\
& 0 \le \epsilon_g \le a \label{eq:minEnonUni2a}
\end{align}
where $\psi=\chi^{-1}$. Note that by Lemma \ref{lem:inverseChi}, $\psi$ is strictly decreasing, convex, and differentiable. Thus, it follows that \eqref{eq:minEnonUni2a}
is a convex optimization  problem \cite{BoydV2004}. 

To lower bound energy consumption, one can solve \eqref{eq:minEnonUni2a}
using a general method like projected gradient descent. However, it would involve projection onto the intersection of $2^n$ half-spaces (in $\mathbb{R}^n$) at each iteration, which is computationally intensive for a large circuit. Here, building on insights from KKT conditions and the structure of  \eqref{eq:minEnonUni2a}, we derive a simple lower-bounding procedure that can be used for any reasonable target $\delta$. This procedure gives simple closed-form expressions for certain structured circuits and also gives rules of thumb for obtaining lower bounds for general circuits. Towards that we make certain observations about problem \eqref{eq:minEnonUni2a}.

First, note that for any $\delta$, there is a finite $\{e_g\}$ that satisfy the constraints. Hence, the value of the objective is finite for this $\{e_g\}$. On the other hand, suppose at the optimum, for some $g$, $\epsilon_g=0$. Then by the definition of physical energy-failure functions we have that $\psi(e_g)=\infty$. This implies an infinite value of the objective. Hence, at the optimum value, $\epsilon_g>0$ for all $g \in V_g$. Hence, problem \eqref{eq:minEnonUni2a} remains the same if we remove the condition $\epsilon_g\ge 0$ for all $g$.

Second, we claim that if the reliability requirement $\delta$ is such that $h(\delta) \le 1 - \exp(-4a)$, then at the optimum of  \eqref{eq:minEnonUni2a}, $\epsilon_g < a$ for all $g \in V_g$. This follows by noting that at 
the optimum, the condition: $\sum_{g \in \mc{P}} \epsilon_g \le \gamma(\delta)$
must be satisfied for all maximal paths $\mc{P}$, and by noting that $\gamma(\delta)=\frac{1}{4}\ln\frac{1}{1-h(\delta)}$.

Hence, for reliability requirement $\delta$ satisfying $h(\delta) \le 1 - \exp(-4a)$, the bound on energy consumption can be obtained by solving:
\begin{align}
& \min \sum_{g \in V_g} \psi(\epsilon_g) \nonumber \\
& \mbox{s.t.}~\sum_{g \in \mc{P}} \epsilon_g \le \gamma(\delta) \mbox{ for all maximal } \mc{P}. 
\label{eq:minEnonUni2}
\end{align}

This problem is also a convex problem satisfying Slater's conditions. So, strong duality holds and the KKT conditions are necessary and sufficient conditions for optimality. In practice, $a$ is at least $0.5$, as without any energy the gate outputs a floating binary value at random, which is correct with a probability at least $0.5$. Thus, the condition is equivalent to $h(\delta) \le 0.865$, which is practically reasonable.

The Lagrangian for problem \eqref{eq:minEnonUni2} is 
\begin{equation}
\sum_{g \in V_g} \psi(\epsilon_g) + \sum_{\mc{P}} \nu_{\mc{P}} \left(\sum_{g \in \mc{P}_i} \epsilon_g - \gamma(\delta)\right), \nonumber
\end{equation}
where $\nu_{\mc{P}} \ge 0$, for all $ \mc{P}$. The KKT conditions are
\begin{align}
& \nu_{\mc{P}} \left(\sum_{g \in \mc{P}} \epsilon_g - \gamma(\delta)\right) = 0~\forall \mc{P}, \nonumber \\
& \sum_{g \in \mc{P}} \epsilon_g \le \gamma(\delta)~\forall \mc{P}, \nonumber \\
& \psi'(\epsilon_g) + \sum_{\mc{P}: g \in \mc{P}} \nu_{\mc{P}} = 0~\forall g. 
\label{eq:KKTminE} 
\end{align}

Now, we use structural properties of the gate graph and the optimization problem to obtain relations among $\epsilon_g$ values at the optimum. Consider any node $g$ in the gate graph and its children $g_1, g_2, \ldots, g_k$. As the gate graph is a tree, any path $\mc{P}_i$ that passes through a $g_l$, $1\le l \le k$, also passes through $g$ and a path that passes through $g_l$ does not pass through $g_{l^{'}}$, $1 \le l\neq l'\le k$. So,
$\sum_{\mc{P}: g \in \mc{P}} \nu_{\mc{P}} = \sum_{l=1}^k \sum_{\mc{P}: g_l \in \mc{P}} \nu_{\mc{P}}$.

This, together with the KKT conditions \eqref{eq:KKTminE} imply that
for any node $g$ in $G_g$ and its children $g_1, g_2, \ldots, g_k$,
\begin{equation}
\psi'(\epsilon_g) = \sum_{l=1}^k \psi'(\epsilon_{g_l}). \label{eq:childSum}
\end{equation}

We make another observation that will be useful later. We claim that at the optimum of
\eqref{eq:minEnonUni2},
\begin{equation}
\sum_{g \in \mc{P}} \epsilon_g = \gamma(\delta),
\label{eq:allPathSame}
\end{equation}
for any maximal path $\mc{P}$.

To see this, let us assume the contrary so there is a path $\mc{P}'$ such 
that $\sum_{g \in \mc{P}'} \epsilon_g < \gamma(\delta)$. As $G_g$ is a tree, there is a leaf node on $\mc{P}'$ through which no other path passes. We can increase
$\epsilon_g$ for that node until the inequality is matched with equality. This would result in a decrease in total energy, which contradicts the optimality of the present energy allocation. So, condition \eqref{eq:allPathSame} holds.

The following lemma is useful in deriving the bound on energy consumption.

\begin{lemma}
\label{lem:condKKT}
Conditions \eqref{eq:childSum} and \eqref{eq:allPathSame} are necessary and sufficient for
$\{\epsilon_g\}$ to be an optimum of \eqref{eq:minEnonUni2}.
\end{lemma}
\begin{IEEEproof}
 As \eqref{eq:minEnonUni2} satisfies Slater's conditions, KKT conditions 
are necessary and sufficient for optimality. So, \eqref{eq:childSum} derived from
KKT conditions is necessary. 
Necessity of \eqref{eq:allPathSame} at the optimum has already been argued.
For proving sufficiency, note that the KKT conditions are sufficient. For each $\mc{P}$,
$\nu_{\mc{P}}=-\psi'(\epsilon_{g_{\mc{P}}})$ where $g_{\mc{P}}$ is the leaf gate
of the path $\mc{P}$. It is easy to check that for these choices of $\nu_{\mc{P}}$,
the KKT conditions are satisfied, given \eqref{eq:childSum} and \eqref{eq:allPathSame} are
satisfied.
\end{IEEEproof}

\subsection{Minimum Energy for Device Technologies}

Let us consider polynomial and
exponential energy-failure functions, and discuss how our optimality conditions yield minimum energy requirements for symmetric gate graphs. Later, we also
present a generic procedure to determine the minimum energy requirement  for any tree gate graph and any physical energy-failure function. 
Immediate derivatives of these procedures for obtaining energy lower-bounds are heuristic schemes for
energy allocation in formulas. 

An exact energy allocation problem for formulas involves finding an energy allocation across gates to minimize total energy consumption while ensuring $\delta$-reliability. Note that the $\delta$-reliability condition requires that the probability of error must be no more than $\delta$ for each input configuration.
Hence, the exact energy allocation problem would have $2^n$ constraints, which are, in general, non-convex. The optimization problem in \eqref{eq:minEnonUni2a} can be seen as a tractable convex surrogate to this problem.

\subsubsection{Polynomial energy-failure function}
\label{sec:polyProcedure}
Consider a polynomial energy-failure function $\epsilon_g = \chi(e_g) = \frac{a}{(1+e_g)^\beta}$. Note that this function has an inverse $e_g=\psi(\epsilon_g)=\left(\frac{a}{\epsilon_g}\right)^{\frac{1}{\beta}}-1$ that is convex, strictly decreasing and differentiable,
$\psi'(\epsilon_g)= - {\frac{1}{\beta}} a^{\frac{1}{\beta}} \epsilon_g^{{\frac{1}{\beta}}-1}$. Hence, condition \eqref{eq:childSum} becomes
\begin{equation}
\epsilon_g^{{\frac{1}{\beta}}-1}=\sum_{l=1}^k \epsilon_{g_l}^{{\frac{1}{\beta}}-1}.\label{eq:childSumPoly}
\end{equation}

Consider a symmetric $k$-ary tree of depth $d$ as the gate graph, i.e., each gate has $k$ inputs. Now, by symmetry  and by condition  \eqref{eq:allPathSame}, each gate at depth $d$ must have the same energy and hence, the same failure probability, say $\epsilon$ at the optimum of \eqref{eq:minEnonUni2}.
Now, again by symmetry and condition  \eqref{eq:allPathSame}, each gate at depth $d-1$ must have the same failure probability, say $\epsilon'$. By condition \eqref{eq:childSumPoly}, we have that
$${\epsilon'}^{{\frac{1}{\beta}}-1}= k \epsilon^{{\frac{1}{\beta}}-1}\mbox{, which implies that }\epsilon'=k^{\frac{\beta}{1-\beta}} \epsilon.$$

Following this procedure implies that at depth $i$ all the gates have the same
failure probability $\epsilon(i)$: 
$$\epsilon(i) = k^{\frac{(d-i)\beta}{1-\beta}} \epsilon.$$

Now, by condition \eqref{eq:allPathSame} it follows that
$\sum_{i=0}^d \epsilon(i) = \gamma(\delta)$,
so we have
$$\epsilon \frac{1-\tilde{k}^{d+1}}{1-\tilde{k}}=\gamma(\delta),$$
where $\tilde{k}=k^{\frac{\beta}{1-\beta}}<1$. This gives the optimum $\epsilon$ to be
$$\epsilon^*=\frac{\gamma(\delta)(1-\tilde{k})}{1-\tilde{k}^{d+1}}.$$

So, the minimum energy requirement is given by the total energy consumed at this optimum allocation:
\begin{align}
\sum_{i=0}^d \psi(\epsilon(i)) k^i 
&= \sum_{i=0}^d k^i (\left(\frac{a}{\epsilon^* \tilde{k}^{d-i}}\right)^{\frac{1}{\beta}}-1) \nonumber \\
&= \sum_{i=0}^d (k\tilde{k}^{\frac{1}{\beta}})^i \tilde{k}^{{\frac{-d}{\beta}}} \left(\frac{a}{\epsilon^*}\right)^{\frac{1}{\beta}} - 
\sum_{i=0}^d k^i \nonumber \\
&= \Omega(n), \nonumber
\end{align}
where the last line follows by noting that $k^{d+1} = \Theta(n)$. 

Observe that in this example of a symmetric circuit with non-uniform energy allocation, a linear scaling of energy with the number of inputs is possible.  This is in contrast to the circuits with uniform operating points, where linear scaling is not possible.
 
Recall from Sec.~\ref{sec:uniform} that any circuit (including ones with reliable components) with $n$ inputs must have energy scaling at least $\Omega(n)$ since for $n$ inputs, we need at least $\frac{n}{k}$ gates with each gate consuming $\Omega(1)$ energy. For reliable components this is obvious, as the energy consumption per gate is constant and does not need to be tuned. For circuits with unreliable components, if there is a gate that consumes $o(1)$ energy, then $\epsilon=\chi(o(1))$ for that gate is $\omega(1)$. Then, the necessary condition $\sum_{g \in \mc{P}} \epsilon_g \le \gamma(\delta)$ cannot be satisfied for any finite $\delta$ for the maximal path through that gate, when $n$ scales.

\begin{remark}
For symmetric gate graphs with polynomial energy-failure functions, we obtain the following rule of thumb for failure allocation: $\epsilon$ should be geometric along the depth with a factor $k^{\frac{\beta}{1-\beta}}$ from a layer to the one above. This in turn gives a rule of thumb for energy allocation: a ratio of $k^{\frac{1}{\beta-1}}$  is maintained from a layer to the one above for  $1+e_g$.
\end{remark}

\subsubsection{Exponential energy-failure function}
\label{sec:expProcedure} 
Now consider the exponential energy-failure function, $\epsilon_g = a \exp(-c e_g)$ with inverse $\psi(\epsilon_g)= - \frac{1}{c} \ln\frac{\epsilon_g}{a}$. Clearly, this function is convex, strictly decreasing, and differentiable,
$\psi'(\epsilon_g)= - \frac{a}{c \epsilon_g}$. Thus, condition \eqref{eq:childSum} becomes
\begin{equation}
\frac{1}{\epsilon_g} = \sum_{l=1}^k \frac{1}{\epsilon_{g_l}}. \label{eq:childSumExp}
\end{equation}

If we consider the same symmetrical $k$-ary tree as the gate graph, for gates at depth $i$ and
$i+1$ we have
$\epsilon(i) = \frac{1}{k} \epsilon(i+1)$,
following the same approach as for polynomial energy-failure functions.
Further following the same steps, we obtain
\begin{align}
\epsilon^* = \frac{\gamma(\delta)(1-\frac{1}{k})}{1-\frac{1}{k^{d+1}}}.
\end{align}

The rest of the derivation of the lower bound follows by replacing $\tilde{k}$ in Sec.~\ref{sec:polyProcedure} by $\frac{1}{k}$ and $\psi(\epsilon_g)=- \frac{1}{c} \ln\frac{\epsilon_g}{a}$. Minimum energy consumption is given by
\begin{align}
\frac{1}{c} \sum_i k^i \ln\frac{a k^{d-i}}{\epsilon^*} 
& = \frac{1}{c} \sum_i \left( (d-i) k^i \ln k + k^i \ln\frac{a(1-\frac{1}{k^{d+1}})}{\gamma(\delta)(1-\frac{1}{k})} \right) = \Omega(n). \nonumber
\end{align}

Notice that the energy allocations for gates at depth
$i$ and $i+1$ follow $\exp(-c (e(i)-e(i+1))) = \frac{1}{k}$,
implying $e(i) = e(i+1) + \frac{1}{c} \ln k$. 

\begin{remark}
For exponential energy-failure function and symmetric gate graph, though the rule of thumb of $\epsilon$ allocation is geometric upwards in the tree (with a factor $k$), the energy allocation is additive with a factor proportional to $\ln k$.
\end{remark}

\subsubsection{Generic Procedure}
\label{sec:genericProcedure}
In general, a numerical value for the minimum energy requirement can be obtained by solving the convex optimization problem \eqref{eq:minEnonUni2} directly for a given $\delta$. This gives an allocation $\{\epsilon_g\}$ and an energy allocation $\{e_g\}$. Note this is only a heuristic energy allocation procedure rather than a provably optimum one, as \eqref{eq:minEnonUni2} only gives necessary conditions to achieve $\delta$-reliability. Though, note no energy allocation scheme can have $\delta$-reliability with a total energy less than the optimum of \eqref{eq:minEnonUni2}.

Alg.~\ref{alg:minE} gives a procedure based on the optimality conditions \eqref{eq:childSum} and \eqref{eq:allPathSame}, which is computationally simpler than solving \eqref{eq:minEnonUni2} using a generic convex optimization algorithm. 

Let us define $\mathfrak{P}($Subtree$, g)$ as the sum of $\epsilon_g$ along any path from
root of the subtree to a leaf through $g$. Let $\mathcal{S}(g)$ be the sibling nodes of
$g$ in the gate graph and subtree$(g)$ be the subtree rooted at $g$.

\begin{algorithm}[ht]
\caption{Minimum Energy Allocation}
\label{alg:minE}
{\bf Input}: $\gamma(\delta)$

{\bf Initialize}: $\forall g, \underline{\epsilon}_g =0, \bar{\epsilon}_g = 1$,
$\gamma=\gamma(\delta)$

{\bf Parameters}: $0<\eta\ll 1$
{\fontsize{10}{10}\selectfont
\begin{algorithmic}[1]
\STATE Call \textsc{SubtreeAlloc}($\gamma(\delta)$,$G_g$).
\STATE Allocate $\psi(\epsilon_g)$ energy to gate $g$
\end{algorithmic}
}
\end{algorithm}

\begin{algorithm}[ht]
\label{alg:SubtreeAlloc}
\caption{\textsc{SubtreeAlloc}($\gamma$,Subtree(root))}
{\fontsize{10}{10}\selectfont
\begin{algorithmic}[1]
\STATE Among nodes at maximum depth pick $g$ with smallest index
\STATE Assign $\epsilon_g=\frac{\underline{\epsilon}_g + \bar{\epsilon}_g}{2}$ to $g$
\WHILE{$|\mathfrak{P}($Subtree$, g)-\gamma|>\eta \gamma$}
\IF{$\mathfrak{P}($Subtree$, g)-\gamma>\gamma \eta$}
\STATE $\bar{\epsilon}_g \leftarrow \epsilon_g$
\ELSE
\STATE $\underline{\epsilon}_g \leftarrow \epsilon_g$
\ENDIF
\STATE Assign $\epsilon_g=\frac{\underline{\epsilon}_g + \bar{\epsilon}_g}{2}$ to $g$
\STATE For $g'$, where $g \in \mbox{child}(g')$: $\epsilon_{g'}\leftarrow \arg\min_{\epsilon} \left(\psi(\epsilon) - \epsilon \sum_{g \in \mbox{child}(g')} \psi'(\epsilon_{g})\right)$ \label{eq:childSumConvex}
\FOR{$v \in$ $\mathcal{S}(g')$ (in numerical order)}
\STATE \hspace{-0.12 in} \textsc{SubtreeAlloc}($\mathfrak{P}($Subtree$(g), g)$),{Subtree($v$)})
\ENDFOR
\ENDWHILE
\STATE For all $u \in \mathcal{S}(g)$,  reset $\underline{\epsilon}_u =0, \bar{\epsilon}_u = 1$, but keep ${\epsilon}_u$
\end{algorithmic}
}
\end{algorithm}

This algorithm starts by allocating some $\epsilon$ to the nodes at lowest depth that are siblings and tunes $\epsilon$ according to a binary search while adjusting $\epsilon$ values of other nodes according to conditions \eqref{eq:childSum} and \eqref{eq:allPathSame}. For any allocations of $\epsilon$ values to the children of a node, it uses condition \eqref{eq:childSum} to obtain the allocation of their parent. Note that solving \eqref{eq:childSum} is the same as solving the one-dimensional unconstrained convex optimization problem in Step \ref{eq:childSumConvex} of \textsc{SubtreeAlloc}. Since $\psi$ is convex, the optimality condition for Step \ref{eq:childSumConvex} is equivalent to the derivative being $0$, which is the same as \eqref{eq:childSum}. We write Step \ref{eq:childSumConvex} in this fashion to show that \eqref{eq:childSum} is easily solvable, given the $\epsilon$ values of child nodes.
When a node has been allocated an $\epsilon$, then one can
enforce conditions on $\epsilon$ values of other sibling nodes using \eqref{eq:allPathSame}. The sub-routine \textsc{SubtreeAlloc} carries out this procedure recursively. 

\begin{proposition}
Alg.~\ref{alg:minE} reaches the optimum of \eqref{eq:minEnonUni2} in $O(|V_g|~Q~\log\frac{1}{\eta})$ steps, where $Q$ is the number of steps required to solve
the one-dimensional convex optimization problem in Step \ref{eq:childSumConvex} of 
\textsc{SubtreeAlloc}.
\end{proposition}
\begin{IEEEproof}
We build on Lemma \ref{lem:condKKT}.
Note that at the optimum of \eqref{eq:minEnonUni2}, conditions \eqref{eq:childSum} and \eqref{eq:allPathSame} are satisfied with path sums being $\gamma(\delta)$. Conditions \eqref{eq:childSum} and \eqref{eq:allPathSame} constrain failure values of all nodes while leaving only one free parameter.
This implies that if the conditions \eqref{eq:childSum} and \eqref{eq:allPathSame} are enforced, then for a given choice of $\epsilon$ for a leaf node at the maximum depth, failure values of all other nodes become fixed (as a function of $\epsilon$). Thus it is then sufficient to do a binary search over $\epsilon$. The algorithm does exactly that. 

At the end of the algorithm, conditions \eqref{eq:childSum} and \eqref{eq:allPathSame} are satisfied with path sum being $\gamma(\delta)$ from any leaf node. As this is  a necessary and sufficient condition for optimality, the algorithm output  is optimal.

The complexity result follows because for a given value of $\epsilon$ on a leaf node, one has to iterate through all nodes and enforce conditions \eqref{eq:childSum} and \eqref{eq:allPathSame}, which takes $O(|V_g|)$ time. In addition, in each iteration, solving the optimization problem in Step \ref{eq:childSumConvex} takes $Q$ queries. As we have to do binary search for $\epsilon$ up to an accuracy $\eta$, the $\log\frac{1}{\eta}$ factor follows.
\end{IEEEproof}

\subsection{Maximum Reliability}
\label{sec:maxReliabilityNonUni}

Let us consider the case where there is a given energy budget $E$ for the whole circuit. Then, the necessary condition for $\delta$-reliability is
\begin{align}
&  \sum_{g \in \mc{P}} \epsilon_g \le \gamma(\delta) \mbox{ for all maximal } \mc{P}, \nonumber \\
& \sum_{g \in G_g} e_g \le E, \nonumber \\
& \epsilon_g = \chi(e_g) \mbox{ and } e_g \ge 0~\forall g, \label{eq:nonUniCond1}
\end{align}

We aim to determine the maximum reliability achievable for a given circuit
energy budget, using the following optimization problem.
\begin{align}
& \min y \nonumber \\
& \mbox{s.t.} \sum_{g \in \mc{P}} \epsilon_g - y \le 0, \mbox{ for all maximal } \mc{P},
\nonumber \\
& \sum_g \psi(\epsilon_g) \le E, \nonumber \\
& 0 \le \epsilon_g \le a~\forall g \in V_g. \label{eq:nonUniCond3a}
\end{align}

Let $y_{min}(E)$ be the solution of problem \eqref{eq:nonUniCond3a}.
Note that $\gamma:[0,\frac{1}{2}]\to[0,\infty)$ is a strictly increasing function of $\delta$ on $[0,\frac{1}{2}]$ and is bijective, with an inverse that is also strictly increasing. This implies $\gamma^{-1}(y_{min}(E))$ is the minimum $\delta$ for which a feasible $\{e_g\}$ exists that satisfies the necessary conditions.

Note that the optimum is strictly decreasing with increasing $E$. To see this, consider $E'>E$, then the optimal solution for budget $E$ is a feasible solution for the problem with budget $E'$. Now use this allocation with additional $\frac{E'-E}{|V_g|}$ energy per gate. This is a feasible solution for the problem with budget $E'$. This feasible solution gives a $y$ value strictly less than $y_{min}(E)$, as $\chi$ is strictly decreasing. Hence, $y_{min}(E')<y_{min}(E)$. Also,  as $E\to \infty$, $y_{min}(E) \to 0$. This is easy to see by picking $\epsilon_g=\frac{E}{|V_g|}$ for all $g$, and noting that $\sum_{g \in \mc{P}} \epsilon_g \le \chi\left(\frac{E}{|V_g|}\right) \max_{\mc{P}}|\mc{P}| \to 0$ as $E \to \infty$. Thus the optimum solution of \eqref{eq:nonUniCond3a}, $y_{min}(E)$ goes to $0$ as $E \to \infty$.

By properties of physical energy-failure functions, namely $\chi(e_g)\to 0$ as $e_g \to \infty$ and $\chi$ strictly decreasing, the conditions $0 \le \epsilon_g$ are redundant for any finite energy budget.
Also, as $y_{min}(E) \to 0$ when $E\to \infty$, there exists $E_{\theta}$ such that for all $E\ge E_{\theta}$, $y_{min}(E)<a$. Thus, for $E\ge E_{\theta}$, $\epsilon_g<a$  for all $g$ are redundant conditions.

There is no closed-form expression for $E_{\theta}$, but is straightforward to compute. For a given budget $E$, let the optimal solution of \eqref{eq:nonUniCond3} be $y^*$; then $E$ is the optimal solution of \eqref{eq:minEnonUni2} for $\gamma(\delta)=y^*$. This is proven by contradiction. Note that the optimal solution of \eqref{eq:nonUniCond3}  is a feasible solution of \eqref{eq:minEnonUni2}. Now, if for a given $\gamma(\delta)=y^*$, the optimal solution of \eqref{eq:minEnonUni2} is strictly less than $E$, then we can divide the excess energy ($E$ minus the minimum) among all gates to improve $y^*$ in \eqref{eq:nonUniCond3}, which is a contradiction. This implies that to compute $E_{\theta}$ we can run
Alg.~\ref{alg:minE} for $\gamma(\delta)=a$.

Now we focus on maximum reliability for an energy budget $E \ge E_{\theta}$, as in most cases $a=0.5$ and $\gamma(\delta)<0.5$ is a bare-minimum reliability
requirement. So, we consider the following problem.
\begin{align}
& \min y \nonumber \\
& \mbox{s.t.} \sum_{g \in \mc{P}} \epsilon_g - y \le 0, \mbox{ for all maximal } \mc{P},
\nonumber \\
& \sum_g \psi(\epsilon_g) \le E. \label{eq:nonUniCond3}
\end{align}

By convexity of $\psi$, this is a convex program. As for
minimum energy allocation, we find KKT conditions for sufficiency of optimality. The conditions are: for $\lambda_{\mc{P}}\ge 0$ and $\mu\ge 0$, 
\begin{align}
& \sum_{\mc{P}} \lambda_{\mc{P}} = 1 \nonumber \\
& \sum_{\mc{P}: g \in \mc{P}_i} \lambda_{\mc{P}} + \mu \psi'(\epsilon_g) = 0,\nonumber\\
&\mu (\sum_{g\in V_g} \psi(\epsilon_g)-E) = 0, \nonumber \\
& \lambda_{\mc{P}} (\sum_{g \in \mc{P}} \epsilon_g - y) = 0 \nonumber.
\end{align}

The following conditions are more useful as they do not involve dual variables.
\begin{equation}
\psi'(\epsilon_g) = \sum_{l=1}^k \psi'(\epsilon_{g_l}). \label{eq:childSumMaxRel}
\end{equation}
\begin{equation}
\sum_{g \in \mc{P}} \epsilon_g = y_{min}(E),
\label{eq:allPathSameMaxRel}
\end{equation}
\begin{equation}
\sum_g \psi(\epsilon_g) = E, \label{eq:matchingBudget}
\end{equation}
for any maximal path $\mc{P}$ from a leaf to the root. 

\begin{lemma}
\label{lem:condKKTReli}
Conditions \eqref{eq:childSumMaxRel}--\eqref{eq:matchingBudget} are necessary and sufficient for the optimality of \eqref{eq:nonUniCond3}.
\end{lemma}
\begin{IEEEproof}
Necessity of \eqref{eq:childSumMaxRel} follows from the necessity of KKT conditions.

Necessity of \eqref{eq:allPathSameMaxRel} follows because at the optimum, path-sums of $\epsilon$ along all maximal paths must match. Otherwise, there is a path $\mc{P}$ for which an $\epsilon$ path-sum is smaller. Then one can take a little energy from the leaf gate in that path and distribute it equally among leaf gates of all other paths. One can choose the amount to be arbitrarily small such that this reallocation of energy: (a) decreases path-sums of all paths other than $\mc{P}$, (b) increases the path-sum of $\mc{P}$, and (c) keeps the path-sum of $\mc{P}$ as the smallest. This strictly decreases $y$, which contradicts the optimality of the current allocation.

Necessity of \eqref{eq:matchingBudget} follows because if there is a strict inequality, we can always distribute the remaining energy to the gates and decrease $\epsilon$ of each gate, resulting in a decrease in $y$. This is contrary to optimality.

For sufficiency of \eqref{eq:childSumMaxRel}--\eqref{eq:matchingBudget}, take $\lambda_{\mc{P}} = \frac{\psi'(\epsilon_{g_{\mc{P}}})}{\sum_{\mc{P}}\psi'(\epsilon_{g_{\mc{P}}})}$, where
$g_{\mc{P}}$ is the leaf of $\mc{P}$ and take $\mu=-\frac{1}{\sum_{\mc{P}}\psi'(\epsilon_{g_{\mc{P}}})}$. Then, it is straightforward to verify these $\lambda_{\mc{P}}$ and $\mu$ values satisfy the KKT conditions. 
\end{IEEEproof}

To derive the best reliability achievable for a given energy budget, we can use the same approach. First, for a given $\epsilon$ at the maximum depth, find $\epsilon$ for all other nodes using \eqref{eq:childSumMaxRel} and \eqref{eq:allPathSameMaxRel}, and then using $\sum_g \chi(\epsilon_g)=E$, obtain $\epsilon$ at the maximum depth. This eventually yields the sum of $\epsilon_g$ along any path, and therefore $y_{min}(E)$.

\subsection{Maximum Reliability for Device Technologies}

\subsubsection{Polynomial energy-failure function}
By the same symmetry arguments as before, gates at the same depth must have the same failure probabilities. Also, by conditions \eqref{eq:childSumMaxRel}--\eqref{eq:allPathSameMaxRel}:
$\epsilon(i) = \tilde{k} \epsilon(i+1)$.
So, if the failure probability at  depth $i$ is $\epsilon(i)$ and the energy budget is $E$, then we must have
\begin{align}
E 
&= \sum_{i=0}^d \psi(\epsilon(i)) k^i \nonumber \\
&= \sum_{i=0}^d k^i \left(\left(\frac{a}{\epsilon(d) \tilde{k}^{d-i}}\right)^{\frac{1}{\beta}}-1 \right) \nonumber \\
&= \sum_{i=0}^d (k\tilde{k}^{\frac{1}{\beta}})^i \tilde{k}^{{\frac{-d}{\beta}}} \left(\frac{a}{\epsilon(d)}\right)^{\frac{1}{\beta}} - 
\sum_{i=0}^d k^i \nonumber \\
&={\epsilon(d)}^{-\frac{1}{\beta}} \sum_{i=0}^d (k\tilde{k}^{\frac{1}{\beta}})^i \tilde{k}^{{\frac{-d}{\beta}}} {a}^{\frac{1}{\beta}} - 
\sum_{i=0}^d k^i, \nonumber
\end{align}
which gives $\epsilon(d)$ in terms of $E$.

The best $\delta$ that can be achieved is
\[
\gamma^{-1}\left(\sum_{i=0}^d \epsilon(i)\right) = \gamma^{-1}\left(\epsilon(d) \frac{1-{\tilde{k}}^{d+1}}{1-\tilde{k}}\right).
\]

\subsubsection{Exponential energy-failure function}
Following the same steps as in the case of polynomial energy-failure functions we obtain a similar expression for $\epsilon(d)$ in terms of $E$, 
\begin{align}
E 
& = \sum_{i=0}^d k^i \frac{1}{c} \ln\frac{a}{\epsilon(i)} \nonumber \\
& = \frac{1}{c} \sum_{i=0}^d k^i \left(\ln\frac{a}{k^{d-i}}-\ln\epsilon(d)\right) \nonumber,
\end{align}
which implies
$$\epsilon(d) = \exp\left(-\frac{c E - \sum_{i=0}^d k^i \ln\frac{a}{k^{d-i}}}{\sum_{i=0}^d k^i}\right)\mbox{ and }\gamma(\delta) = \epsilon(d) \frac{1-\frac{1}{k^{d+1}}}{1-\frac{1}{k}}.$$

\subsubsection{General procedure}
A general procedure of energy allocation can be obtained by solving convex program \eqref{eq:nonUniCond3}. As in minimum energy allocation, a computationally simpler approach would be to use the necessary and sufficient KKT conditions. We, however, design Alg.~\ref{alg:maxReli} to use minimum energy allocation as a sub-routine. Similar correctness and complexity results hold with an additional multiplicative factor of $\log\frac{1}{\theta}$ in the computational complexity, where $\theta$ is the energy budget accuracy $(1\pm\theta)E$.

The following lemma is useful in proving the accuracy of Alg.~\ref{alg:maxReli}.
\begin{lemma}
\label{lem:minEmaxReliEqv}
For a given energy budget $E$, $y_{min}$ is the optimum of \eqref{eq:nonUniCond3} if and only if for a given reliability requirement $\gamma(\delta)=y_{min}$, $E$ is the optimum of
\eqref{eq:minEnonUni2}. 
\end{lemma}
\begin{IEEEproof}
As argued before, recall that for a physical energy-failure function, $y_{min}(E)$ is a strictly decreasing function of $E$, $y_{min}:[0,\infty) \mapsto [0,\infty]$, which is one-to-one and onto. 

We start with the direct part. As $E$ meets $\gamma(\delta)$ reliability requirement
in \eqref{eq:minEnonUni2} for  an allocation of $\epsilon_g$, in \eqref{eq:nonUniCond3}
this $\epsilon_g$ allocation meets the budget $E$ while achieving a $\gamma(\delta)$ reliability. So, $y_{min}(E) \le \gamma(\delta)$. Let $y_{min}(E)<\gamma(\delta)$, then we can take away some energy from some gates in the circuit, so that their $\epsilon_g$ increases and
$\sum_{g \in \mc{P}} \epsilon_g= \gamma(\delta)$ for all maximal $\mc{P}$. This implies that
the total energy used is less than $E$ while achieving $\gamma(\delta)$ reliability, which is a contradiction.

The converse follows similarly by starting with an allocation that achieves maximum reliability $y_{min}(E)$ for a given budget $E$ and showing $E$ is the minimum energy requirement for a given reliability criteria $\gamma(\delta)=y_{min}(E)$. Proof follows similarly by showing an inequality, and then showing that a strict inequality is a contradiction.
\end{IEEEproof}

Based on Lemma~\ref{lem:minEmaxReliEqv}, in Alg.~\ref{alg:maxReli} we do a binary search over $\delta$
to find the minimum $\delta$ for which minimum energy allocation obtained is $E$.

\begin{algorithm}[ht]
\caption{Maximum Reliability Allocation}
\label{alg:maxReli}

{\bf Input}: $E$

{\bf Initialize}: $\forall g, \underline{\epsilon}_g =0, \bar{\epsilon}_g = 1$,
$\gamma=\gamma(\delta)$

{\bf Parameters}: $0<\eta\ll 1$, $0<\theta\ll 1$
{\fontsize{10}{10}\selectfont
\begin{algorithmic}[1]
\STATE $\underline{\delta} =0, \bar{\delta} = 1$, $\delta=\frac{1}{2}$.
\WHILE{$|\sum_g \psi(\epsilon_g) - E|>\theta E$}
\IF{$\sum_g \psi(\epsilon_g) - E > \theta E$}
\STATE $\underline{\delta} \leftarrow \delta$
\ELSE
\STATE $\bar{\delta} \leftarrow \delta$
\ENDIF
\STATE $\delta=\frac{\underline{\delta}+\bar{\delta}}{2}$
\STATE Call \textsc{SubtreeAlloc}($\gamma(\delta)$,$G_g$)
\ENDWHILE
\STATE Allocate $\psi(\epsilon_g)$ energy to gate $g$
\end{algorithmic}
}
\end{algorithm}

\begin{proposition}
\label{prop:maxReli}
Alg.~\ref{alg:maxReli} reaches the optimum of \eqref{eq:nonUniCond3} in $O(|V_g|~Q~\log\frac{1}{\eta})$ steps for a given $\theta$ if $E\ge E_{\theta}$, where $Q$ is the number of steps required to solve
the one-dimensional convex optimization problem in Step \ref{eq:childSumConvex} of
\textsc{SubtreeAlloc}.
\end{proposition}

\begin{IEEEproof}
There is a unique $\delta$ for an energy budget by Lemma \ref{lem:minEmaxReliEqv}, so it is sufficient to search over all $\delta$ to find a $\delta$ for which $E$ is the minimum
energy requirement. Hence, doing a binary search for $\delta$ and running the minimum energy procedure for each $\delta$ converges to the correct reliability requirement.

Complexity guarantees follow from similar results for the minimum energy procedure. Here we perform an additional binary search for $\delta$, but since it is over $[0,1]$, this search takes $O(1)$ iterations which is included in the constants in the complexity order result.
\end{IEEEproof}

Note that the energy allocation rules in this section are derived from necessary conditions for $\delta$-reliability. Hence, they are heuristics for achieving $\delta$-reliability. But, as they are derived based on the necessary conditions, no energy allocation scheme can achieve a better energy consumption (for a given reliability) or reliability (for a given energy budget) than the bounds obtained by these allocation schemes.

\section{Feedforward Neural Networks}
\label{sec:dnn}

Now we extend the mutual information propagation technique from tree-structured circuits to directed acyclic graphs, to study binary feedforward neural networks.  Recall the basic problem formulation from Sec.~\ref{sec:nndef}.  Though we restrict to single-output neural networks, most results in the sequel can be extended to neural networks with multiple outputs by appropriately modifying the definition of $\delta$-reliability.

We give a preliminary lemma before pursuing more general insight.
Consider the setting of neural networks where neurons in a given layer
have the same operating point.
\begin{lemma}
\label{lem:NNinfoProp}
For an $L$-layer neural network with $n$ inputs, a single output, and the same failure probabilities for all neurons in a layer:
\[
1-h(\delta) \le \sum_{P \in \mc{P}_{g}} \prod_{g \in P} (1-2\epsilon_g)^2, 
\]
for all $g \in \mc{N}_1$ (the input layer), where $\mc{P}_{g}$ is the set of directed paths from $g \in \mc{N}_1 $ to the
output.
\end{lemma}
\begin{IEEEproof}
Proof follows using the proof of a similar result for a circuit with uniform $\epsilon$ in \cite{Evans1994}.
\end{IEEEproof}

As neurons in the same layer must have the same energy, let  $e^{(\ell)}$ 
denote the energy of a neuron in layer $\ell$ and let the failure probability of a
neuron in layer $\ell$ be denoted $\epsilon^{(\ell)}=\chi(e^{(\ell)})$. Then, by simple substitution, the condition becomes
\begin{align}
1-h(\delta) \le |\mc{P}_{g}| \prod_{\ell=1}^L (1-2\epsilon^{(\ell)})^2, \label{eq:neceReliCondProd}
\end{align}
for all $g \in \mc{N}_1$. The condition can also be written as
\begin{align}
1-h(\delta) \le \left(\min_{g \in \mc{N}_1} |\mc{P}_{g}|\right)  \prod_{\ell=1}^L (1-2\epsilon^{(\ell)})^2 \mbox{.}
\end{align}

Taking natural logarithms of both sides and using the inequality $1-x\le \exp(-x)$, 
the necessary condition for $\delta$-reliability  becomes:
\begin{align}
\ln\frac{1}{1-h(\delta)} \ge 4 \sum_l \epsilon^{(\ell)} - \min_{g \in \mc{N}_1} \ln|\mc{P}_{g}| \mbox{.}
\label{eq:neceReliCond}
\end{align}

\subsection{Homogeneous Neurons}
\label{sec:uniform}
As discussed in Sec.~\ref{sec:intro}, there are two hardware-determined technological possibilities for energy allocation to neurons: either all neurons must be operated at the same energy level or they can be operated at different energy levels. In this section,
we consider the setting where all neurons are operated at the same energy level. We want to understand the implication of a prescribed $\delta$-reliability of the neural network on its energy consumption in the regime where $L$ is large, i.e.\ the neural network is \emph{deep}.

For an $L$-layer neural network with an underlying directed graph $G$, let us define $\pi^G(L)=\min_{g \in \mc{N}_1} |\mc{P}_g|$. From
\eqref{eq:neceReliCondProd}, for uniform neuron operating points, we have:
\begin{align}
1-h(\delta) \le \pi^G(L) (1-2\epsilon)^{2L}. \label{eq:UnifReliCond}
\end{align}

The total energy consumed by the network is $\psi(\epsilon) \sum_l N_{\ell}$. Hence, for a neural network with uniform energy allocation, a universal lower bound for energy consumption is $O(\sum_{\ell} N_{\ell})$.  

Note that there is also linear scaling in the absence of failures. Thus our universal lower bound is order-optimal, in the sense of scaling with the size of the network up to a constant independent of network size and depth.  To achieve this order-optimal scaling
of neural network energy consumption, neuron failure probability $\epsilon$ must be a constant independent of $L$ and $\sum_{\ell} N_{\ell}$.

If $\epsilon$ is a constant then $(1-2\epsilon)^{2L}$ goes to $0$ exponentially with $L$. Hence, to meet any non-trivial ($\delta<\frac 1 2$) $\delta$-reliability constraint, we must have
\begin{align}
\pi^G(L)\ge \frac{1-h(\delta)}{(1-2\epsilon)^{2L}}.
\end{align}
So, a necessary condition for order-optimal (linear) energy scaling with a per-neuron energy consumption $e_G$, i.e., 
for energy consumption scaling as $e_G \sum_{\ell} N_{\ell}$, a necessary condition is that
$\pi^G(L)$ must grow exponentially with $L$ with an exponent $d_G=1/(1-2\chi(e_g))^2$.
This poses the following constraint on the structure of the deep neural network. 
There must be exponentially many paths, $d_G^L$, from each neuron at the input layer to the
output neuron. This implies that if connectivity between any two of the $L-2 \approx L$ adjacent hidden layers are similar and they are regular, then each hidden neuron must have a degree $d_G$ connectivity to both of its adjacent layers. 
\begin{remark}
Consider a deep neural network of faulty neurons with uniform connectivity across all adjacent hidden layers.  Then a linear (in per layer size) number of edges between any two adjacent layers is necessary for reliability
of the network.
\end{remark}

To gain further insights on structural properties of reliable neural networks constructed from faulty neurons, let us temporarily restrict attention to the class of uniformly-noisy neural networks with $L$ layers and $r$ neurons in each hidden layer, such that all hidden neurons have $d$ directed edges to the next layer. We want to understand the minimum energy required for networks in the class to be able to compute $\delta$-reliably, where the minimum is taken over \emph{all} $n$-input boolean functions from $\{0,1\}^n$ to $\{0,1\}$ that can be computed using this class of neural networks.

Since there are $r$ neurons in each hidden layer and the error probability of each neuron is $\epsilon$, then for large $L$ the total energy consumed by the network is
$(L-2)~r \cdot\psi(\epsilon)\approx L~r \cdot\psi(\epsilon)$.

To obtain the minimum energy bound, we compute the minimum of this quantity
 subject to the necessary $\delta$-reliability constraint in \eqref{eq:UnifReliCond}.
Replacing $\pi_G(L)$ by $d^L$ yields a lower bound on the minimum
energy consumption. This enumeration arises from the maximum number of paths possible from an input
neuron to the output neuron if all the adjacent hidden layers are fully connected, which
gives $d^L$ paths. 

To satisfy \eqref{eq:UnifReliCond}, we need $\epsilon\le \frac{1}{2}(1-1/\sqrt{d})$, as
for large $L$, for any non-trivial constant $\delta$, the term $\left(\ln \frac{1}{1-h(\delta)}\right)^{\frac 1 L}$
is $1$. So, the minimum energy consumed within the class of uniform and regular neural networks with $L$ layers scales as 
$\approx L ~ r\cdot\psi\left( \frac{1}{2}\left(1-\frac{1}{\sqrt{d}}\right)\right)$.

Hence, the minimum energy consumption lower bound is linear in the number of hidden 
layers and the scaling constant depends on the particular energy-failure function, connectivity between
layers, and the number of neurons in each layer.

Consider the exponential energy failure function $\epsilon_g = a \exp(-c e_g)$, then 
$\psi(\epsilon)=\frac{1}{c} \ln\frac{a}{\epsilon}$. In this case the energy scaling is linear with
the size of the network $(\approx L\cdot r$) and the constant is
$\frac{1}{c} \ln \frac{2 a \sqrt{d}}{\sqrt{d}-1}$.

\subsection{Heterogeneous Neurons}
\label{sec:nonUniform}
Next we investigate the setting with flexible hardware technologies, with neurons that may have  different energy operating points. In particular, consider the case where neurons in the same layer have the same energy operating points, but the operating points may differ across layers. The total energy consumption of this network is
$\sum_{\ell} N_{\ell} \psi(\epsilon^{(\ell)})$.
We aim for a lower bound on the
total energy needed for $\delta$-reliability.

As \eqref{eq:neceReliCond} is a necessary condition for $\delta$-reliability,
the solution to the following optimization problem gives an energy lower bound. 
\begin{align}
& \min_{\epsilon^{(\ell)},1\le \ell\le L} \sum_{\ell} N_{\ell} \psi(\epsilon^{(\ell)}) \nonumber \\
& \mbox{s.t.} \sum_{\ell} \epsilon^{(\ell)} \le \gamma(\delta), \nonumber \\
& \  0\le \epsilon^{(\ell)}\le 1~\mbox{ for all } \ell, \label{eq:nonUniOpti}
\end{align}
where $\gamma(\delta)=\frac{1}{4}\left(\ln\frac{1}{1-h(\delta)}+\ln\pi^G(L)\right)$. 

For any energy-failure function $\chi$ for which $\lim_{\epsilon \to 0}\chi^{-1}(\epsilon)=\infty$,
the constraint on non-negativity of $\epsilon^{(\ell)}$ is redundant.
So the Lagrangian of this problem is:
$$\sum_{\ell} N_{\ell} \psi(\epsilon^{(\ell)}) + \lambda (\sum_{\ell} \epsilon^{(\ell)}-\gamma(\delta))+\sum_{\ell} \nu_{\ell} \epsilon^{(\ell)}
-\sum_{\ell} \nu_{\ell},$$
for $\lambda \ge 0$ and $\nu_{\ell} \ge 0$ for all $\ell$.

KKT conditions for this problem are:
\begin{align}
N_{\ell} \psi'(\epsilon^{(\ell)}) &= -\lambda-\nu_{\ell} \mbox{ for all } \ell, \nonumber \\
\lambda \left(\sum_{\ell} \epsilon^{(\ell)}-\gamma\right) &= 0, \nonumber \\
\nu_{\ell}(\epsilon-1)&=0, \nu_{\ell} \ge 0 \mbox{ for all } \ell, \nonumber \\
\sum_{\ell} \epsilon^{(\ell)} &\le \gamma, \nonumber \\
\lambda &\ge 0. \nonumber
\end{align}

First we argue that at optimum $\sum_{\ell} \epsilon^{(\ell)} = \gamma$. Otherwise, one can always
increase some $\epsilon^{(\ell)}$ to meet the inequality with equality and also simultaneously 
reduce the energy consumption. This implies that $\lambda$ is not 
constrained to be $0$. Now, as $\psi$ is strictly decreasing, $\psi'>0$ for all
$\epsilon$, and hence $\lambda>0$. So, we have
\begin{align}
N_1 \psi'(\epsilon^{(1)}) + \nu_1 &= N_2 \psi'(\epsilon^{(2)}) + \nu_2 \notag \\ 
&~~ \vdots  \notag  \\
&= N_{L} \psi'(\epsilon^{(L)}) + \nu_L. \label{eq:condNu} 
\end{align}

If optimization problem \eqref{eq:nonUniOpti} has parameter $\gamma(\delta)$, which depends on $\delta$ and $G$, and an energy-failure function $\chi$ such that $\epsilon^{(\ell)}<1$ for all $\ell$, then there is a simple relation at the
optimum that can be used to compute $\{\epsilon^{(\ell)}\}$:
\begin{align} 
N_1 \psi'(\epsilon^{(1)}) = N_2 \psi'(\epsilon^{(2)}) = \cdots = N_{L} 
\psi'(\epsilon^{(L)}), \label{eq:desRule}
\end{align}
as $\nu_{\ell}=0$ for all $\ell$ under this condition.

Note that if we remove the constraints $\epsilon^{(\ell)}\le 1$ from \eqref{eq:nonUniOpti} and obtain an optimum
in $[0,1)^L$ for that relaxed problem, then that is also the optimum of \eqref{eq:nonUniOpti}. Hence, if we use the
relation in \eqref{eq:desRule} and obtain a solution in $[0,1)^L$, then that is the optimum solution of
\eqref{eq:nonUniOpti}. 

Let us consider some specific energy-failure functions.  For a polynomial energy-failure function $\chi(e_g) = a/(1+e_g)^\beta$, $\beta>1$,
under $G$ and $\delta$ for which \eqref{eq:desRule} is the optimality condition,
we have
$ N_{\ell} (\epsilon^{(\ell)})^{\frac{1}{\beta}-1} = N_{\ell+1} (\epsilon^{(\ell+1)})^{\frac{1}{\beta}-1}$, which implies
$$\epsilon^{(\ell+1)}=\epsilon^{(\ell)} \left(\frac{N_{\ell}}{N_{\ell+1}}\right)^{\frac{\beta}{1-\beta}}.$$

Similarly, for  exponential energy-failure function $\chi(e_g)=a \exp(-c e_g)$,
the condition is
$N_{\ell}/ \epsilon^{(\ell)} = N_{\ell+1}/\epsilon^{(\ell+1)}$, which implies
$\epsilon^{(\ell+1)}=\frac{N_{\ell+1}}{N_{\ell}}\epsilon^{(\ell)}$.

In these cases, the minimum energy consumption bound follows by choosing a variable $\epsilon$ 
for layer $L$, which is the output layer and has one neuron.
Then following the relation in \eqref{eq:desRule}, $\epsilon^{(\ell)}$, $\ell \le L-1$ can be obtained in terms of 
$\epsilon$ and can solve
the equation $\sum_{\ell} \epsilon^{(\ell)} = \gamma$ to obtain $\epsilon$. After substituting the
value of $\epsilon$ and those of $\epsilon^{(\ell)}$, obtained from the value of $\epsilon$, into
the objective of \eqref{eq:nonUniOpti} the energy bound follows.

Consider an exponential energy-failure function, first assuming
at optimum $\epsilon^{(\ell)}<1$, i.e., $\nu_{\ell}=0$ for all $\ell$. Let $\epsilon^{(\ell)}=\epsilon$, then
$\epsilon^{(\ell)}=\epsilon N_{\ell}$, as $N_{\ell}=1$ and hence,
$ \epsilon(1+\sum_{l \le L-1} N_{\ell})=\gamma$,
which implies 
$\epsilon = \gamma/\sum_{\ell} N_{\ell}$.
The minimum energy bound for exponential energy-failure function is therefore:
\begin{align}
& \frac{1}{c} \sum_{\ell} N_{\ell} \ln\frac{a\sum_{\ell} N_{\ell}}{\gamma N_{\ell}} \nonumber \\
&\quad = \frac{1}{c} \sum_{\ell} N_{\ell} \ln\frac{\sum_{\ell} N_{\ell}}{N_{\ell}} + \frac{1}{c} \ln\frac{a}{\gamma} \sum_{\ell} N_{\ell} \nonumber \\
&\quad = \frac{1}{c} \sum_{\ell} N_{\ell} \ln\frac{N}{N_{\ell}} + \frac{1}{c} \ln\frac{a}{\gamma} \left(\sum_{\ell} N_{\ell}\right),
\nonumber
\end{align}
where $N$ is the total number of neurons. For a non-trivial constant $\delta<1/2$, $\gamma$ is
almost independent of $\delta$ if $\pi^G(L)$ grows sufficiently fast with $L$. In that case the energy lower bound
depends only on the structure of the neural network, reemphasizing the importance of the neural graph structure in reliability.

Let us restrict attention to the class of deep neural networks made of regularly connected hidden layers as in Sec.~\ref{sec:uniform}.
We obtain a lower bound on the energy consumption for non-uniform operating points across the layers. 
To compute a closed-form lower bound on energy, we need to use the simple relation among $\{\epsilon^{(\ell)}\}$
that has been derived above under the condition that the optimum is in $[0,1)^L$. 
For this class of networks,
$N_{\ell} = r$ and $\sum_{\ell} N_{\ell} = (L-1)r+1$. Hence, $\epsilon = \frac{\gamma}{(L-1)r+1}$ and 
$\epsilon^{(\ell)}\approx\frac{\gamma}{L-1}$ for large $L$. Note that if
$\gamma < L-1$ then the obtained $\{\epsilon^{(\ell)}\}$ are  in $[0,1)^L$ and hence are the
optima of \eqref{eq:nonUniOpti}. 

Note that if $\ln d < 4$, then for large $L$ we have $\gamma < L-1$ and hence $\{\epsilon^{(\ell)}\}$ solutions obtained
from \eqref{eq:desRule} are optimal for \eqref{eq:nonUniOpti}. So, the energy lower bound is
\[
\frac{1}{c} N \ln L - \frac{1}{c} N (\ln L + \ln\ln d) + \frac{1}{c} N \ln 4a  = \frac{1}{c} N \ln \frac{4a}{\ln d}.
\]
As $\ln d < 4$, for $a=1$, i.e.\ the neuron surely fails  at zero energy, the energy lower bound scales linearly in
the size of the network, and the scaling constant is 
$\frac{1}{c} \ln \frac{4}{\ln d}$. By comparison with
the bound in Sec.~\ref{sec:uniform} for uniform operating point and the same energy-failure function,
it follows that the energy bound for non-uniform operating points scales linearly with
number of neurons with a strictly smaller constant.

Conditions on $G$ and $\gamma$ under which \eqref{eq:condNu} is equivalent to \eqref{eq:desRule} also imply
that the energy (or failure probability) allocations at the optimum of \eqref{eq:nonUniOpti} have a simple relation across layers. As \eqref{eq:neceReliCond} is a necessary condition for $\delta$-reliability and not a sufficient condition, the solution of
\eqref{eq:nonUniOpti} cannot be proved to achieve $\delta$-reliability. Nevertheless, the solution gives a feasible energy
allocation of the neural network which can serve as a design heuristic.

\section{Numerical Examples and Practical Insights}
\label{sec:numericals}
In previous sections,  we obtained a heuristic for energy allocation en route to bounding reliability. In this section we numerically characterize the performance of the heuristics in simple settings. As this work is primarily concerned with obtaining fundamental energy-reliability bounds, a detailed study of energy allocation methods is out of scope.  Complementary work details practical applications in circuit design \cite{ShanbhagVKPV2019}.

For the purpose of our simple numerical studies, we consider three different boolean functions widely used in information theory and computer science: conjunction, disjunction, and parity. We study circuits constructed with {\small AND} gates for 
conjunction, circuits constructed with {\small OR} gates for disjunction, and circuits constructed with {\small XOR} gates for parity. In all cases we consider $2$-input 
gates with exponential energy-failure functions, the fundamental thermodynamic bound for CMOS devices \cite{Kish2004}.

\begin{figure}
\centering
\subfigure[Tree]{
\includegraphics[scale=0.15]{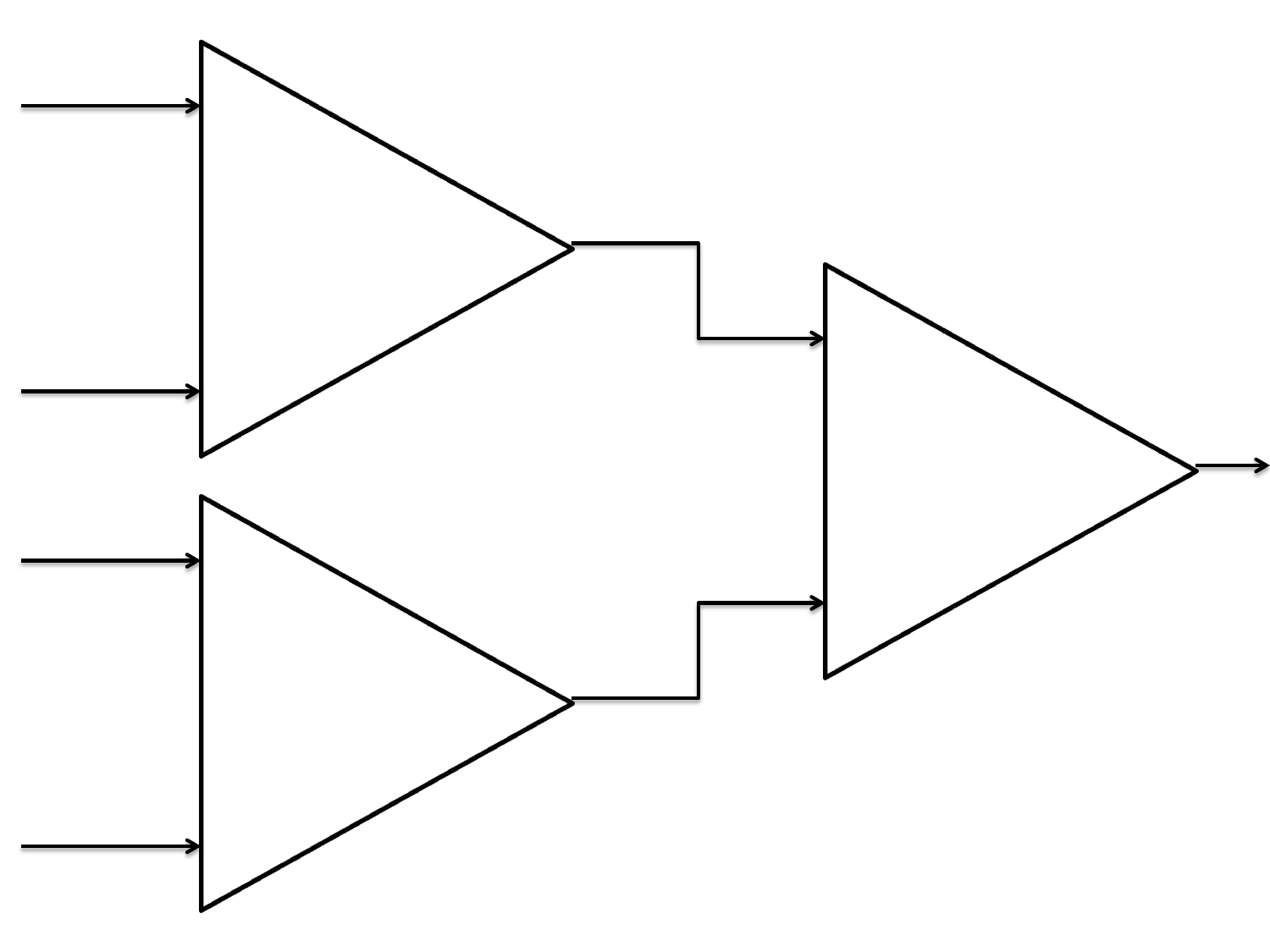}
\label{fig:tree}
}
\subfigure[Line]{
\includegraphics[scale=0.15]{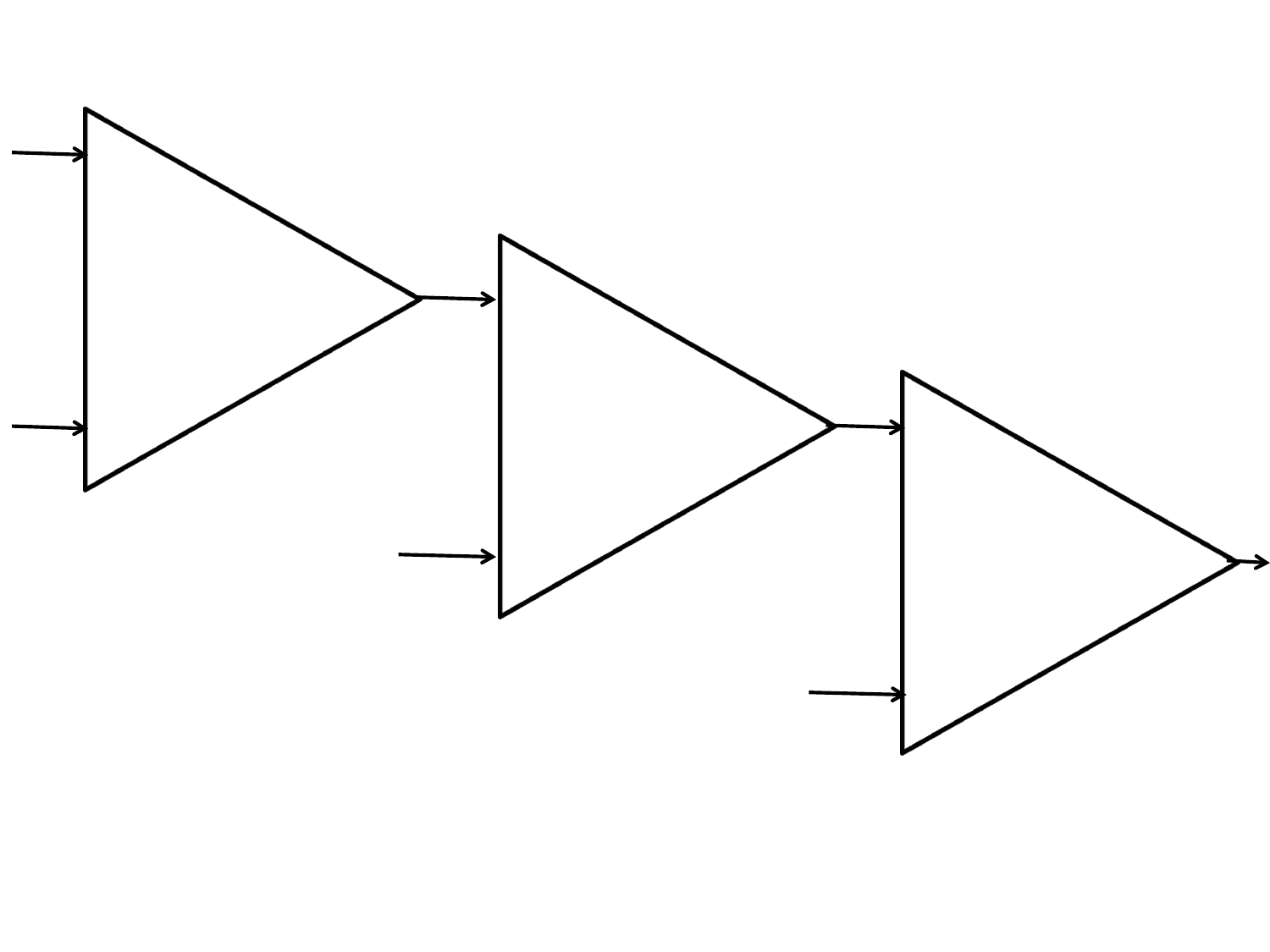}
\label{fig:line}
}
\caption{Tree and line circuit structures.}
\label{fig:graph}
\end{figure}

A conjunction circuit is constructed with {\small AND} gates in two main ways: line graph and symmetric tree, as in Fig.~\ref{fig:graph}. The same is true for disjunction circuit constructed with {\small OR} gates and parity circuits constructed with {\small XOR} gates. We study reliability of both configurations for each of these three types of circuits, under the heuristic energy allocation schemes. Though the energy allocation scheme is applicable to any $n$-input boolean function, in this section we restrict ourselves to $4$-input boolean functions. Small circuits are insightful since we can compute and precisely evaluate closed-form expressions for reliability.

For the exponential energy-failure function $\epsilon = \epsilon_0 \exp(-c e_g)$, in most cases $\epsilon_0=0.5$, as the output is perfectly random when no energy is allocated to the gate. Here we study the reliability of the circuits for a given energy budget, $E \ge \sum_g e_g$. For convenience we specify the budget in terms of $c E$ such that $c E \ge \sum_g (c e_g) = \sum_g \ln\frac{\epsilon_0}{\epsilon_g}$. 

\begin{figure}
\centering
\subfigure[Energy allocated to individual gates.]{
\includegraphics[scale=0.45]{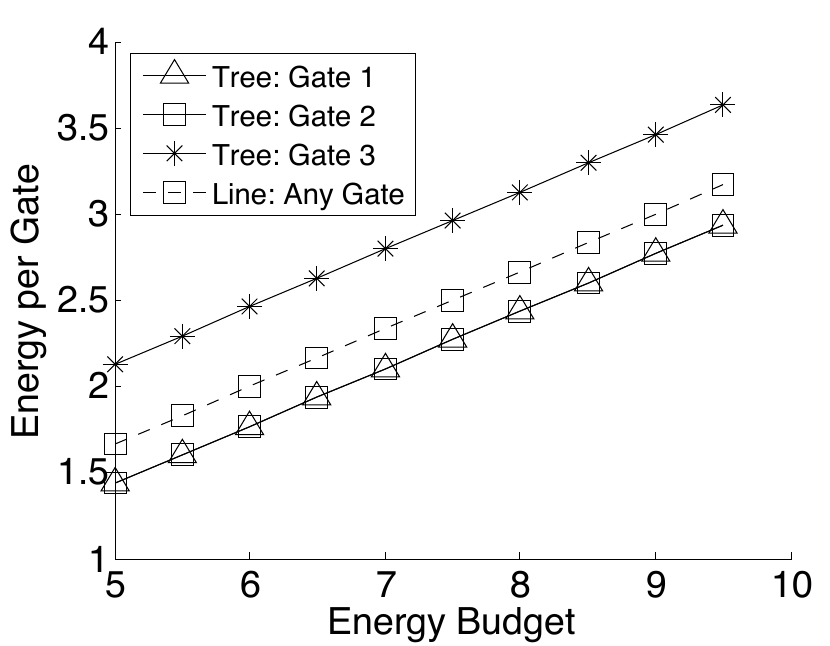}
\label{fig:energyAlloc}
}
\subfigure[Error entropy limit for worst-case inputs.]{
\includegraphics[scale=0.45]{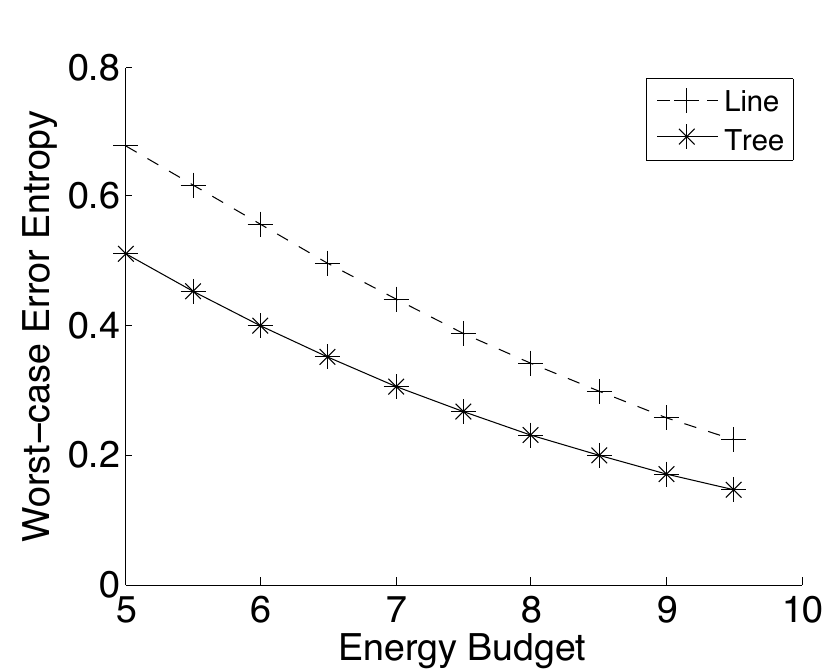}
\label{fig:targetEntropy}
}
\caption{Energy allocation and worst-case error entropy.}
\end{figure}

Note that though the heuristic depends on the circuit structure, it is not affected by the types of gates. Hence, the heuristic energy allocation is same for conjunction, disjunction, and parity. For the line circuit, it is not hard to see that the heuristic allocation rule gives uniform energy allocation. On the other hand, for tree circuits, the heuristic allocation gives different energies to the gates. In Fig.~\ref{fig:energyAlloc} we plot the energy allocated to each gate ($c e_g$) for different energy budgets ($c E$). Note that under the heuristic scheme, the output gate (gate 3) is allocated more energy to increase its reliability.

As discussed before, $\delta$-reliability is equivalent to  worst-case (across all input patterns) error entropy, $h(\delta)$ (for $\delta\le 0.5$). Fig.~\ref{fig:targetEntropy} plots the limit of worst-case error entropy against different energy budgets. We observe that the tree graph has a better worst-case entropy bound than the line graph.

Next, we study the individual boolean functions separately. The performance measure we choose is conditional error entropy, $H(E|X_1, X_2, \ldots, X_n)$, where $E$ is the binary error variable and $\{X_i\}$ are binary input variables, as this also captures the effects of inputs on the error entropy.
For each boolean function, we compute closed-form expressions for $H(E|X_1, X_2, \ldots, X_n)$ for both tree and line graphs in terms of $\epsilon_g$ for each gate. Then, we evaluate these omitted expressions for each energy (and hence $\epsilon_g$) allocation.  For each boolean function, we study conditional error entropy for line graph under the heuristic allocation, for tree under the heuristic allocation, and for tree under uniform energy allocation. Note that the heuristic allocation is uniform for line graph.

\begin{figure}
\centering
\subfigure[Conjunction.]{
\includegraphics[scale=0.45]{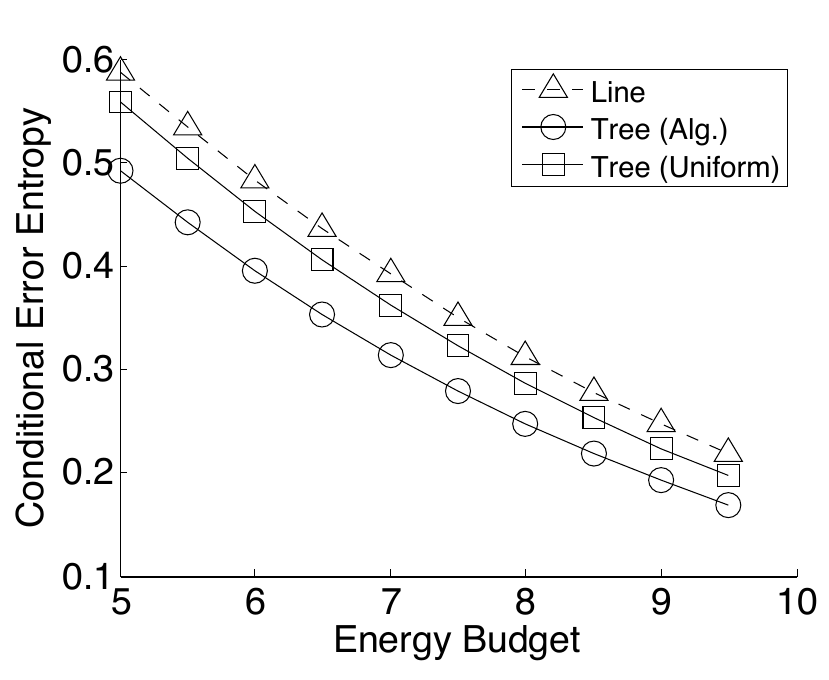}
\label{fig:ANDperf}
}
\subfigure[Disjunction.]{
\includegraphics[scale=0.45]{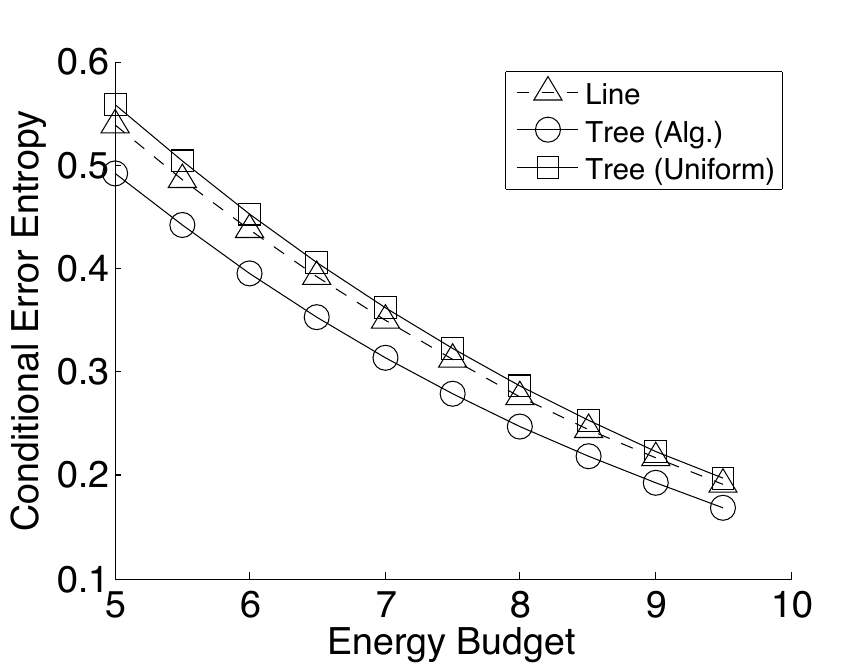}
\label{fig:ORperf}
}
\subfigure[Parity.]{
\includegraphics[scale=0.45]{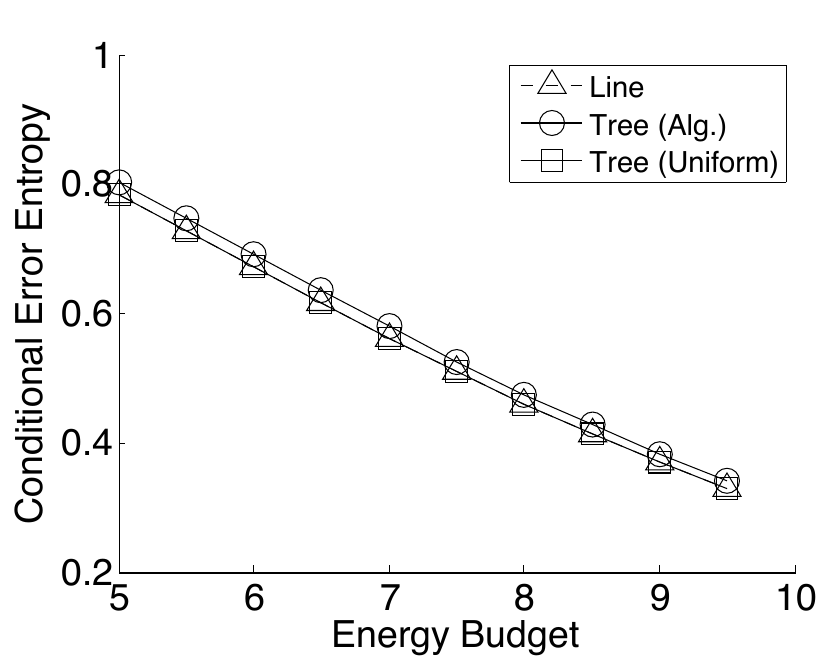}
\label{fig:XORperf}
}
\caption{Conditional error entropy performance under heuristic allocation.}
\end{figure}

Fig.~\ref{fig:ANDperf} plots the conditional error entropy again energy budget for conjunction. We observe that the tree graph with heuristic energy allocation performs better than line graph with heuristic energy allocation (uniform allocation). We also observe that tree graph with uniform energy allocation has a better error performance than line graph. For conjunction, a tree graph is more reliable than a line graph. 

But the same argument is not true for disjunction. In Fig.~\ref{fig:ORperf} we see that tree graph with uniform energy allocation has a worse error performance than line graph. But, tree graph with heuristic energy allocation performs better than line graph with heuristic energy allocation. In both conjunction and disjunction circuits we observe that the heuristic energy allocation for tree graphs perform better (by 15--18\%) than uniform energy allocation. 

But, this does not mean that the heuristic allocation is uniformly best. One shortcoming of this heuristic energy allocation is that it optimizes a necessary condition for $\delta$-reliability while ignoring the truth table of individual gates. This is reflected in case of parity circuits in Fig.~\ref{fig:XORperf}. Notice that the error probabilities are the same for both tree and line graphs. If we consider each gate to be a gate plus an independent Bernoulli noise, then error happens if an odd number of gates have Bernoulli noise $1$. This is true irrespective of circuit structure. Thus all gates are equally critical for reliable realization of this boolean function and uniform energy allocation is optimal.
In summary, the energy allocation heuristic that emerges from bounding fundamental energy-reliability limits may or may not be effective. 

To study the implications of our energy allocation insights in the neural network setting, in work presented elsewhere \cite{ChatterjeeV2019}, we made predictions about mammalian sensory cortex, under the optimization approach to theoretical biology.  Experimentally-testable hypotheses were consistent with experimental observations, in the sense that different aspects of neural connectivity, reliability, and energy are all matched to one another as predicted.

\section{Conclusion and Future Work}
\label{sec:discussion}
Given deep neural networks at the application layer and nanoscale devices at the physical layer are both emerging technologies, there is a desire to implement one on the other for on-device inference.  In pursuing this vision, we need to understand the effect nanoscale device unreliability has on the energy and performance of neural networks. Our scaling bounds for energy
consumption led to insights into the structural and connectivity requirements of reliable neural networks and also offered design heuristics. 

As part of this investigation, we obtained a lower bound on the minimum energy needed to compute an $n$-input boolean function using unreliable gates. We observed that a superlinear scaling of energy with the number of input bits is unavoidable for sub-exponential energy-failure functions, irrespective of the reliability requirement, when gates are constrained to have uniform operating points. Contrarily, when gates are allowed to have different operating points, minimum energy requirements for polynomial and exponential energy-failure functions and symmetric circuits demonstrate that the lower bound is linear in the number of inputs.  This argues for the value of emerging device technologies that allow variable gate operations. For general circuits and energy-failure functions, we proposed an algorithm that can numerically compute the lower bound in linear time. This procedure also gives a heuristic to allocate energy to different gates. The heuristic energy allocation is generic irrespective of the constituent gate types and optimizes the necessary conditions for $\delta$-reliability. The scheme is computationally simple and may perform well in several settings.

As future work, we aim to develop provably optimal energy allocation schemes with uniform performance guarantees for general circuits.
This work has followed the worst-case reliability paradigm of von Neumann for 
logic circuits, but future work aims to investigate average-case reliability requirements.

\section*{Acknowledgment}
Discussions with Ameya Patil and Naresh R. Shanbhag are appreciated.

\bibliographystyle{IEEEtran}
\bibliography{conf_abrv,abrv,lrv_lib}

\newcommand{\SortNoop}[1]{}
\begin{thebibliography}{10}
\providecommand{\url}[1]{#1}
\csname url@samestyle\endcsname
\providecommand{\newblock}{\relax}
\providecommand{\bibinfo}[2]{#2}
\providecommand{\BIBentrySTDinterwordspacing}{\spaceskip=0pt\relax}
\providecommand{\BIBentryALTinterwordstretchfactor}{4}
\providecommand{\BIBentryALTinterwordspacing}{\spaceskip=\fontdimen2\font plus
\BIBentryALTinterwordstretchfactor\fontdimen3\font minus
  \fontdimen4\font\relax}
\providecommand{\BIBforeignlanguage}[2]{{%
\expandafter\ifx\csname l@#1\endcsname\relax
\typeout{** WARNING: IEEEtran.bst: No hyphenation pattern has been}%
\typeout{** loaded for the language `#1'. Using the pattern for}%
\typeout{** the default language instead.}%
\else
\language=\csname l@#1\endcsname
\fi
#2}}
\providecommand{\BIBdecl}{\relax}
\BIBdecl

\bibitem{ChatterjeeV2016}
A.~Chatterjee and L.~R. Varshney, ``Energy-reliability limits in nanoscale
  circuits,'' in \emph{Proc. 2016 Inf. Theory Appl. Workshop}, Feb. 2016.

\bibitem{ChatterjeeV2017a}
------, ``Energy-reliability limits in nanoscale neural networks,'' in
  \emph{Proc. 51st Annu. Conf. Inf. Sci. Syst. (CISS 2017)}, Mar. 2017.

\bibitem{SchwartzDSE2019_arXiv}
R.~Schwartz, J.~Dodge, N.~A. Smith, and O.~Etzioni, ``Green {AI},'' Jul. 2019,
  arXiv:1907.10597 [cs.CY].

\bibitem{KangKSEC2014}
M.~Kang, M.-S. Keel, N.~R. Shanbhag, S.~Eilert, and K.~Curewitz, ``An
  energy-efficient {VLSI} architecture for pattern recognition via deep
  embedding of computation in {SRAM},'' in \emph{Proc. IEEE Int. Conf. Acoust.,
  Speech, Signal Process. (ICASSP 2014)}, May 2014, pp. 8326--8330.

\bibitem{WangLV2015}
Z.~Wang, K.~H. Lee, and N.~Verma, ``Overcoming computational errors in sensing
  platforms through embedded machine-learning kernels,'' \emph{{IEEE} Trans.
  {VLSI} Syst.}, vol.~23, no.~8, pp. 1459--1470, Aug. 2015.

\bibitem{ZhangS2015}
S.~Zhang and N.~R. Shanbhag, ``Reduced overhead error compensation for energy
  efficient machine learning kernels,'' in \emph{Proc. 2015 IEEE/ACM Int. Conf.
  Comput.-Aided Des. (ICCAD)}, Nov. 2015, pp. 15--21.

\bibitem{AhnHYMC2015}
J.~Ahn, S.~Hong, S.~Yoo, O.~Mutlu, and K.~Choi, ``A scalable
  processing-in-memory accelerator for parallel graph processing,'' in
  \emph{Proc. 42nd Annu. Int. Symp. Comput. Architecture (ISCA '15)}, Jun.
  2015, pp. 105--117.

\bibitem{ShanbhagVKPV2019}
N.~R. Shanbhag, N.~Verma, Y.~Kim, A.~D. Patil, and L.~R. Varshney,
  ``Shannon-inspired statistical computing for the nanoscale era,'' \emph{Proc.
  {IEEE}}, vol. 107, no.~1, pp. 90--107, Jan. 2019.

\bibitem{Neftci2016}
E.~Neftci, ``Stochastic neuromorphic learning machines for weakly labeled
  data,'' in \emph{Proc. IEEE 34th Int. Conf. Comput. Design (ICCD)}, Oct.
  2016, pp. 670--673.

\bibitem{Shanbhag2016_arXiv}
N.~R. Shanbhag, ``Energy-efficient machine learning in silicon: A
  communications-inspired approach,'' arXiv:1611.03109 [cs.LG]., Oct. 2016.

\bibitem{ButlerMMVRRH2012}
W.~H. Butler, T.~Mewes, C.~K.~A. Mewes, P.~B. Visscher, W.~H. Rippard, S.~E.
  Russek, and R.~Heindl, ``Switching distributions for perpendicular
  spin-torque devices within the macrospin approximation,'' \emph{{IEEE} Trans.
  Magn.}, vol.~48, no.~12, pp. 4684--4700, Dec. 2012.

\bibitem{ShanbhagMVOMRJR2008}
N.~R. Shanbhag, S.~Mitra, G.~de~Veciana, M.~Orshansky, R.~Marculescu,
  J.~Roychowdhury, D.~Jones, and J.~M. Rabaey, ``The search for alternative
  computational paradigms,'' \emph{{IEEE} Des. Test. Comput.}, vol.~25, no.~4,
  pp. 334--343, July-Aug. 2008.

\bibitem{GuptaADDGKMNRSSS2013}
P.~Gupta, Y.~Agarwal, L.~Dolecek, N.~Dutt, R.~K. Gupta, R.~Kumar, S.~Mitra,
  A.~Nicolau, T.~S. Rosing, M.~B. Srivastava, S.~Swanson, and D.~Sylvester,
  ``Underdesigned and opportunistic computing in presence of hardware
  variability,'' \emph{{IEEE} Trans. Comput.-Aided Design Integr. Circuits
  Syst.}, vol.~32, no.~1, pp. 8--23, Jan. 2013.

\bibitem{NahlusKSB2014}
I.~Nahlus, E.~P. Kim, N.~R. Shanbhag, and D.~Blaauw, ``Energy-efficient dot
  product computation using a switched analog circuit architecture,'' in
  \emph{Proc. 2014 IEEE/ACM Int. Symp. on Low Power Electronics and Design
  (ISLPED 2014)}, Aug. 2014.

\bibitem{De2016}
V.~De, ``Energy-efficient computing in nanoscale {CMOS},'' \emph{{IEEE} Des.
  Test}, vol.~33, no.~2, pp. 68--75, Apr. 2016.

\bibitem{VonNeumann1956}
J.~von Neumann, ``Probabilistic logics and the synthesis of reliable organisms
  from unreliable components,'' in \emph{Automata Studies}, C.~E. Shannon and
  J.~McCarthy, Eds.\hskip 1em plus 0.5em minus 0.4em\relax Princeton: Princeton
  University Press, 1956, pp. 43--98.

\bibitem{Kish2004}
L.~B. Kish, ``Moore's law and the energy requirement of computing versus
  performance,'' \emph{IEE Proceedings - Circuits, Devices and Systems}, vol.
  151, no.~2, pp. 190--194, Apr. 2004.

\bibitem{DobrushinO1977}
R.~L. Dobrushin and S.~I. Ortyukov, ``Lower bound for the redundancy of
  self-correcting arrangements of unreliable functional elements,''
  \emph{Probl. Inf. Transm.}, vol.~13, no.~1, pp. 82--89, Jan.-Mar. 1977.

\bibitem{PippengerST1991}
N.~Pippenger, G.~D. Stamoulis, and J.~N. Tsitsiklis, ``On a lower bound for the
  redundancy of reliable networks with noisy gates,'' \emph{{IEEE} Trans. Inf.
  Theory}, vol.~37, no.~3, pp. 639--643, May 1991.

\bibitem{GacsG1994}
P.~G{\'{a}}cs and A.~G{\'{a}}l, ``Lower bounds for the complexity of reliable
  {B}oolean circuits with noisy gates,'' \emph{{IEEE} Trans. Inf. Theory},
  vol.~40, no.~2, pp. 579--583, Mar. 1994.

\bibitem{HajekW1991}
B.~Hajek and T.~Weller, ``On the maximum tolerable noise for reliable
  computation by formulas,'' \emph{{IEEE} Trans. Inf. Theory}, vol.~37, no.~2,
  pp. 388--391, Mar. 1991.

\bibitem{Pippenger1988}
N.~Pippenger, ``Reliable computation by formulas in the presence of noise,''
  \emph{{IEEE} Trans. Inf. Theory}, vol.~34, no.~2, pp. 194--197, Mar. 1988.

\bibitem{EvansS1998}
W.~Evans and N.~Pippenger, ``On the maximum tolerable noise for reliable
  computation by formulas,'' \emph{{IEEE} Trans. Inf. Theory}, vol.~44, no.~3,
  pp. 1299--1305, May 1998.

\bibitem{Evans1994}
W.~S. Evans, ``Information theory and noisy computation,'' Ph.D. dissertation,
  University of California, Berkeley, Berkeley, CA, 1994.

\bibitem{EvansS1999}
W.~S. Evans and L.~J. Schulman, ``Signal propagation and noisy circuits,''
  \emph{{IEEE} Trans. Inf. Theory}, vol.~45, no.~7, pp. 2367--2373, Nov. 1999.

\bibitem{ChatterjeeV2019}
A.~Chatterjee and L.~R. Varshney, ``Optimal energy allocation in reliable
  neural sensory processing,'' in \emph{Encyclopedia of Computational
  Neuroscience}, D.~Jaeger and R.~Jung, Eds.\hskip 1em plus 0.5em minus
  0.4em\relax New York, NY, USA: Springer, 2019.

\bibitem{AntoniadisBNPS2014}
A.~Antoniadis, N.~Barcelo, M.~Nugent, K.~Pruhs, and M.~Scquizzato,
  ``Energy-efficient circuit design,'' in \emph{Proc. 5th Conf. Innov. Theor.
  Comput. Sci. (ITCS '14)}, Jan. 2014, pp. 303--312.

\bibitem{Pippenger1985}
N.~Pippenger, ``On networks of noisy gates,'' in \emph{Proc. 26th Annu. Symp.
  Found. Comput. Sci.}, Oct. 1985, pp. 30--38.

\bibitem{Karnaugh1953}
M.~Karnaugh, ``The map method for synthesis of combinational logic circuits,''
  \emph{Trans. {AIEE}}, vol.~72, no.~5, pp. 593--599, Nov. 1953.

\bibitem{RussellN2010}
S.~Russell and P.~Norvig, \emph{Artificial Intelligence: A Modern Approach},
  3rd~ed.\hskip 1em plus 0.5em minus 0.4em\relax Upper Saddle River, NJ:
  Prentice Hall, 2010.

\bibitem{BoydV2004}
S.~Boyd and L.~Vandenberghe, \emph{Convex Optimization}.\hskip 1em plus 0.5em
  minus 0.4em\relax Cambridge: Cambridge University Press, 2004.

\end{thebibliography}

\appendix

\begin{IEEEproof}[Proof of Lemma \ref{lem:inverseChi}]
First, we use standard arguments from real analysis to establish $\chi$ has an inverse that is one-to-one and strictly decreasing. Since $\chi$ is strictly decreasing, it is one-to-one. Otherwise there would be an $e_1 < e_2$, such that $\chi(e_1)=\chi(e_2)$. 
From differentiability and therefore continuity of $\chi$, it also follows that $\chi$ is onto. 

Let $0<\epsilon<a$ be such that there is no $e_g>0$ with $\chi(e_g)=\epsilon$ and we show a contradiction.
As the function is decreasing, there exists an $e$ such that $\chi(x) < \epsilon$ for all $x>e_g$ and $\chi(x)>\epsilon$ for all $x<e_g$. Consider
$\lim_{x \uparrow e_g} \chi(x)$ and $\lim_{x \downarrow e_g} \chi(x)$. For any sequence $\{x_k\} \uparrow e_g$, $\{\chi(x_k)\}$ is monotonic and bounded, so $\lim_k \chi(x_k)$ exists and is finite. Similarly, this is true for $x_k \downarrow e_g$ and $\lim_k \chi(x_k)$. As $\chi$ is continuous these two limits match and must equal $\chi(e_g)$. But as $\chi(x)<\epsilon<\chi(x')$ for $x>e_g>x'$, the only possible limit is $\epsilon$. So, $\chi(e_g)=\epsilon$. This implies that $\chi:(0,\infty)\to(0,a)$ is onto.
So, an inverse exists for one-to-one and onto $\chi$.

To see that $\chi^{-1}$ is strictly decreasing, let  $0 < \epsilon_1 < \epsilon_2 < a$. Now, there exists $e_1$ and $e_2$ such that $e_i=\chi^{-1}(\epsilon_i), i \in \{1,2\}$. We show that $e_1\le e_2$ is not possible. Let $e_1\le e_2$, then by the strictly decreasing property of $\chi$, $\chi(e_1) \ge \chi(e_2)$, implying $\epsilon_1 \ge \epsilon_2$, which is contradictory.

Note that the derivative of $\chi^{-1}$ at a value $\epsilon$ is $1$ by the derivative of $\chi$ at the value $\chi^{-1}(\epsilon)$, if it is defined. As $\chi$ is strictly decreasing, at no point the derivative of $\chi$ is $0$, and hence, $\chi^{-1}$ is differentiable.

Without loss of generality let $e_1>e_2$, so $\epsilon_1=\chi(e_1)< \epsilon_2\chi(e_2)$, 
then for $\alpha \in [0,1]$,
\begin{align}
\chi^{-1}(\alpha \epsilon_1 + (1-\alpha)\epsilon_2) 
& =\chi^{-1}(\alpha\chi(e_1)+(1-\alpha) \chi(e_2)) \nonumber \\
&\stackrel{(a)}{\le} \chi^{-1}(\chi(\alpha e_1 + (1-\alpha )e_2)) \\
& = \alpha e_1 + (1-\alpha )e_2 \nonumber \\
& = \alpha \chi^{-1}(\epsilon_1) + (1-\alpha)\chi^{-1}(\epsilon_2), \nonumber
\end{align}
where (a) follows because $\chi^{-1}$ is decreasing and $\chi$ is convex.
This proves convexity of $\chi^{-1}$.
\end{IEEEproof}

\begin{IEEEproof}[Proof of Lemma \ref{lem:leafSize}]
A directed rooted tree is a structure where directed lines originate from leaves and are eventually merged at the root node. Each non-leaf node merges lines coming from the level below it. Thus there are $L$ lines to be merged.

As each node can have at most $k$ children, each non-leaf node can merge at most $k$ lines. Each non-leaf node can be thought of as a gadget/entity which can take at most $k$ lines as input and outputs one merged line. Consider any structure of the non-leaves that merges $L$ lines to $1$. In any of these structures if there is a non-leaf node with less than $k$ inputs, then replacing that with a non-leaf node with $k$ inputs does not increase the total number of non-leaf nodes. So for any tree there is a $k$-ary tree with the same number of leaf nodes, and no more non-leaf nodes.

Note that converting a $k$-ary tree to a balanced $k$-ary tree with the same number of leaf nodes does not increase the total number of nodes and hence, does not increase the number of non-leaf nodes.
\end{IEEEproof}

\begin{IEEEproof}[Proof of Lemma \ref{lem:leafDepth}]
By similar arguments as in Lemma \ref{lem:leafSize}. Note that if there is a tree with certain number of leaves and each node with at most $k$ children, then depth does not increase if we change it to a $k$-ary tree to same number of leaves. The rest follows by noting hat converting a $k$-ary tree to a balanced $k$-ary tree with the same number of leaves does not increase the depth.
\end{IEEEproof}

\begin{IEEEproof}[Proof of Lemma \ref{lem:leafMonotone}]
Consider two directed trees with $L$ and $L'$ leaves, $L'\ge L$. Then, any graph configuration with $L'$ leaves can be transformed to graph with $L$ leaves without increasing the size (just by dropping $L'-L$ leaves). Hence, the minimum-achieving configuration for $L'$ leaves is also a configuration for $L$ leaves. Hence, the minimum for $L$ leaves is not larger than the minimum for $L'$ leaves.
\end{IEEEproof}

\end{document}